\documentclass[twocolumn]{aastex63}
\usepackage{physics}
\usepackage{dsfont}
\usepackage{CJK}
\def\arraystretch{1.5}
\newcommand*{\rowstyle}[1]{
 \gdef\@rowstyle{#1}%
 \@rowstyle\ignorespaces%
}
\received{April 12, 2021}
\accepted{September 2, 2021}
\shorttitle{Universal disk size under non-ideal MHD}
\shortauthors{Lee et al.}

\begin{document}
\begin{CJK*}{UTF8}{bkai}

\title{Universal protoplanetary disk size under complete non-ideal magnetohydrodynamics: \\
The interplay between ion-neutral friction, Hall effect, and the Ohmic dissipation  }

\correspondingauthor{Yueh-Ning Lee}
\email{ynlee@ntnu.edu.tw}

\author[0000-0003-3497-2329]{Yueh-Ning Lee (李悅寧)}
\affiliation{Department of Earth Sciences, National Taiwan Normal University, Taipei 116, Taiwan}
\affiliation{Center of Astronomy and Gravitation, National Taiwan Normal University, Taipei 116, Taiwan}
\affiliation{Physics Division, National Center for Theoretical Sciences, Taipei 106, Taiwan}

\author[0000-0002-4577-8292]{Pierre Marchand}
\affiliation{American Museum of Natural History, Department of Astrophysics, CPW at 79th, New-York, NY, 10024}

\author{Yu-Hsuan Liu (劉宇軒)}
\affiliation{Department of Physics, National Taiwan Normal University, Taipei 116, Taiwan}

\author[0000-0002-0472-7202]{Patrick Hennebelle}
\affiliation{Universit\'{e} Paris Diderot, AIM, Sorbonne Paris Cit\'{e}, CEA, CNRS, F-91191 Gif-sur-Yvette, France}
\affiliation{IRFU, CEA, Universit\'{e} Paris-Saclay, F-91191 Gif-sur-Yvette, France}
\affiliation{LERMA (UMR CNRS 8112), Ecole Normale Sup\'{e}rieure, 75231 Paris Cedex, France}



\begin{abstract}
The role of non-ideal magnetohydrodynamics has been proven critical during the formation of the protoplanetary disk, particularly in regulating its size. We provide a simple model to predict the disk size under the interplay among the ambipolar diffusion, the Hall effect, and the Ohmic dissipation. The model predicts a small disk size of around 20 AU, that depends only sub-linearly on disk parameters, for a wide range of initial conditions of sub-Solar mass and moderate magnetization. It is able to explain phenomena manifested in existing numerical simulations, including the bimodal disk behavior under parallel and anti-parallel alignment between the rotation and magnetic field. In the parallel configuration, the disk size decreases and eventually disappears. In the anti-parallel configuration, and the disk has an outer partition (or pseudo-disk) that is flat, shrinking , and short-lived, as well as a inner partition that grows slowly with mass and is long-lived. Even with significant initial magnetization, the vertical field in the disk can only dominate at the early stage when the mass is low, and the toroidal field eventually dominates in all disks.
\end{abstract}



\section{Introduction} \label{sec:intro}
During the collapse of a prestellar core, the formation of a rotationally supported disk (RSD) is naturally expected as a result of angular momentum (AM) conservation. The effect of the magnetic braking (MB) and the non-ideal magnetohydrodynamics (NIMHD), both significant at this scale, have only started to be taken into account in the last few decades.
The AM conservation in the absence of MB results in a rapidly growing disk that is prone to instabilities, while the MB alone removes the AM too efficiently and hinders the formation of a disk. This phenomenon has been historically referred to as the magnetic catastrophe. The consideration of the NIMHD effects alleviates this catastrophe and the disk becomes somehow self-regulated.

Several recent numerical simulations have implemented the NIMHD effects, including the Hall effect \citep{Tsukamoto2015a,Marchand2018,Zhao2020a,Zhao2020b}, the ambipolar diffusion \citep[AD, or ion-neutral friction;][]{Wurster2014,Tsukamoto2015b,Masson2016,Zhao2020b}, and the Ohmic dissipation \citep{Inutsuka2010,Tsukamoto2015a,Tsukamoto2015b,Zhao2020a,Zhao2020b}. 
\citet{Tsukamoto2015a} found a disk size bimodality in the presence of Hall effect: a small disk of a few AU forms when the magnetic field is parallel to the rotation axis and a disk of $\sim 20$ AU forms when the two vectors are anti-parallel. They refer to the disk formed in these two configurations as ortho-disk and para-disk, respectively. Unlike the diffusion terms, the Hall effect depends on the sign of the dominant charged species, and thus the relative orientation between the rotation and the magnetic field can have an impact on the dynamics. \citet{Zhao2020a} further suggested that the ortho-disk is short-lived, and the para-disk has two components: an inner disk and an outer partition that is more extended and flattened, both in close-to-Keplerian rotation. 
Recent observations also suggest that the class 0/I disks, which are embedded in the collapsing envelope, are likely small and magnetized \citep[$\lesssim 50$ AU][]{Andrews2018,Andrews2020,Maury2018,Maury2019}.
\citet{Hennebelle2016} proposed a model for the self-regulated disk size in the presence of AD. The current work is an extension of this model accounting for all NIMHD effects.

\section{Non-ideal MHD in a collapsing core}\label{sec:nimhd}
Throughout this work, cylindrical coordinate system will be used, where $\hat{r}$, $\hat{\phi}$, and $\hat{z}$ represent radial, azimuthal, and vertical unit vectors. Axisymmetry is assumed for simplicity.

The equation of momentum conservation writes
\begin{eqnarray}\label{eq:momentum}
{\partial (\rho \vb*{u}) \over \partial t} = -\nabla \cdot\left[ \rho \vb*{u}\vb*{u} + \!\left( \!P \!+\! {B^2 \over 2} \right) \!\mathds{I} \!-  \vb*{B}\vb*{B} \right] -\rho \nabla \Phi,
\end{eqnarray}
where $\rho$ is the density, $\vb*{u}$ is the velocity, $P$ is the pressure, $\vb*{B}$ is the magnetic field, and $\Phi$ is the gravitational potential.

The induction equation writes
\begin{eqnarray}\label{eq:B}
{\partial \vb*{B} \over \partial t} &=& \nabla \times (\vb*{u} \times \vb*{B} ) - \nabla \times \left\{ 
 \eta_{\rm A} {\vb*{B} \over \| \vb*{B} \|} \times \left(\nabla \times \vb*{B}  \right) \times {\vb*{B} \over \| \vb*{B} \|}  \right.   \nonumber
\\
&+& \left. \eta_{\rm H} \left(\nabla \times\vb*{B}\right) \times {\vb*{B} \over \|\vb*{B}\|}
+ \eta_{\rm O} \nabla \times \vb*{B} \right\} \\
&=& \nabla \times \left[(\vb*{u}+\vb*{u}_{\rm A}+\vb*{u}_{\rm H}) \times \vb*{B} -\eta_{\rm O} \nabla \times \vb*{B}\right] \nonumber\\
&=& \nabla \times \left(\vb*{v} \times \vb*{B} - \eta_{\rm O} \nabla \times \vb*{B} \right),\nonumber
\end{eqnarray}
where $\eta_{\rm A}$, $\eta_{\rm H}$, and $\eta_{\rm O}$ are the resistivities of the ambipolar diffusion, Hall effect, and Ohmic dissipation, respectively.
The non-ideal MHD effects become dynamically important when the coupling between charged and neutral particles is not perfect, and thus the magnetic field lines move with respect to the fluid motion. 
The ambipolar diffusion and the Hall effect can be expressed as velocities of the field lines ($\vb*{u}_{\rm A}$ and $\vb*{u}_{\rm H}$, of which only the component perpendicular to the magnetic field is effective). The Ohmic dissipation behaves very much the same as the ambipolar diffusion, since the friction is always in the direction of the relative velocity between the field lines and the gas (See Appendix \ref{ap:NIMHD} for more details of the non-ideal MHD equations).

\subsection{Initial conditions}
The discussions are limited to the simple scenario where both the initial magnetic field and the rotation axis are aligned along $z$-axis. 
For a magnetized, rotating core in collapse, the initial condition is characterized by two dimensionless parameters:
the normalized mass-to-flux ratio,
\begin{eqnarray}\label{eq:lambda}
\lambda = {2\pi\sqrt{G}M  \over  \pi r^2 B_z}
\end{eqnarray}
and the rotation-to-gravitational energy ratio,
\begin{eqnarray}\label{eq:beta}
\beta ={ E_{\rm rot} \over E_{\rm grav}}.
\end{eqnarray}

Neglecting the turbulent motion, the two parameters completely determine the status of the core: magnetically supported for small $\lambda$ and rotationally supported for large $\beta$. The thermal energy is comparably small at the core scale and is not considered here. During the collapse of the core, the level of $\beta$ determines whether the disk inherits the initial rotation or is subject to the rotation generated by the azimuthal Hall drift, while the level of $\lambda$ sets the relative strength between $B_\phi$, $B_r$, and $B_z$. The collapsing cores will globally follow the behavior described below.

\subsection{The collapsing envelope}
The envelope is globally collapsing with weak, or zero, rotation, and thus $u_r \approx u_z \gg u_\phi$. 
Initially, $B_z \gg B_r, B_\phi$.
The Hall drift is dominated by its azimuthal component. 
We summarize the global evolution of the system. 
\begin{itemize}
    \item $B_z$ grows by radial advection, and is slowly lost through radial diffusion.
    \item $B_r$ grows by the bending of magnetic field lines from $B_z$, depending on the magnetic criticality of the core, and is lost through diffusion.
    \item $B_\phi$ grows by the rotation drag or Hall induction from $B_z$, depending on the relative strength of the initial rotation compared to the Hall drift velocity. The growth from Hall induction is less rapid than that of $B_r$.
    \item $u_\phi$ is amplified by the collapse if initially non-zero, and the dragged field lines brake the rotation. The azimuthal Hall induction decreases/increases the braking in the para-/ortho-case. If the Hall effect is strong enough (large $B_r$ from collapse), it is possible that the rotation direction is reversed for the parallel (ortho) case. 
\end{itemize}
This behavior is described in \citet{Zhao2020a}, where the rotation is accelerated in the opposite direction of the Hall drift in response to the distorted magnetic field lines. $B_z$ and $B_r$ are lost through the same mechanism, either by the magnetic pressure gradient due to accumulated $B_z$ or by the magnetic tensor that straightens the field lines bent in $r$-direction, depending on the relative strength of the two terms at each location (see Appendix \ref{ap:Br} for detailed discussion of the radial transport).

\subsection{The magnetized pseudo-disk, or the outer disk}
This is a loosely defined region, where a high-density disk-like structure is formed, while not necessarily in Keplerian rotation. 
This corresponds roughly to a rotating region where the field lines are significantly pinched and $B_\phi$ has grown to a level such that $u_{{\rm H},r}$ becomes important. \citet{Zhao2020a} refers to this as the rotationally-supported Hall current sheet (RSHCS). 
\begin{itemize}
    \item $B_z$ continues to grow through advection while the loss through radial diffusion increases. The radial Hall drift also becomes significant as $B_\phi$ grows.
    \item $B_r$ induction diminishes, since $|u_\phi|$ quickly becomes larger than $|u_r|$. 
    $B_r$ is dynamically less important than $B_\phi$.
    \item $B_\phi$ increases inwards due to the collapse that amplifies the rotation, and is transported by the radial Hall drift. The azimuthal drift due to Hall effect is less important here, and thus the direction of the rotation axis is fixed. 
    \item $u_\phi$ also increases inwards due to the collapse. The MB is somehow important such that the rotation does not necessarily reach the Keplerian value.
\end{itemize}

\subsection{Inside the disk}
The velocity approaches the Keplerian rotation and the MB is exactly compensated by the radial accretion of AM. 
\begin{itemize}
    \item $B_z$ is gradually lost through radial Hall drift or diffusion. An equilibrium is only reached when $B_\phi$ decreases to a level where $u_{{\rm H},r}$ becomes comparable to the small value of $u_r$ within the disk.
    \item $B_r$ is dynamically unimportant, and its loss is similar to that of $B_z$.
    \item $B_\phi$ is generated through induction from $B_z$ and lost through azimuthal/vertical diffusion.
    \item $u_\phi$ is almost Keplerian and is less influenced by the magnetic field dynamically. 
\end{itemize}

\section{Theoretical estimate of the radius}\label{sec:model}
Following the same reasoning as in \citet{Hennebelle2016}, we search for an equilibrium radius at the disk-envelope boundary in quasi-stationary state. 
This boundary should correspond to the radius at which dynamical effects of the magnetic field become just sufficient to brake the pseudo-disk that feeds AM to the disk, but will not be able to grow any stronger.
This means that 
   (1) the growth of the azimuthal magnetic field diminishes, and that
   (2) the increase of rotation velocity when crossing the envelope-disk boundary is canceled out by the magnetic breaking.
These two conditions can be translated into comparing the relevant timescales.

\subsection{Properties of the envelope and the disk}
Following \citet{Hennebelle2020b}, 
the density in the envelope is expressed as
\begin{eqnarray}\label{eq:rho_e}
\rho_{\rm e}(r) = {\delta_\rho C_{\rm s}^2 \over 2 \pi G r^2 },
\end{eqnarray}
where $C_{\rm s} = 200~ {\rm m/s}$ is the thermal sound speed, $G$ is the gravitational constant, and the numerical factor $\delta_\rho \gtrsim 1$.
The density right inside the envelope-disk boundary writes
\begin{eqnarray}\label{eq:rho}
\rho(r) = {\delta_\rho C_{\rm s}^2 \over 2 \pi G r^2 } \mathcal{M}^2
= {\delta_\rho C_{\rm s}^2 \over 2 \pi G r^2 } \left(u_r(r)\over C_{\rm s}\right)^2,
\end{eqnarray}
where $\mathcal{M}$ is the Mach number of the accretion shock. 
This formula describes the singular isothermal sphere profile  \citep[SIS,][]{Shu1977} amplified by the accretion shock at the disk boundary.
This is slightly different from what was proposed in \citet{Hennebelle2016}, while the effective behavior is similar. 
The Keplerian velocity is given by
\begin{eqnarray}\label{eq:v_kep}
v_{\rm kep} \simeq \sqrt{GM \over r} = \sqrt{G(M_\ast + M_{\rm d}) \over r},
\end{eqnarray}
where $M$ is the total mass of the system, with $M_\ast$ and $M_{\rm d}$ representing the mass of the star and that of the disk. 
The rotational velocity, $u_\phi = \delta_\phi v_{\rm kep}$, and infall velocity, $u_r = \delta_r v_{\rm kep}$, are both prescribed with $\delta_\phi, \delta_r \lesssim 1$.
Vertical hydrostatic equilibrium is assumed in the neighborhood of the disk, such that the disk scale height
\begin{eqnarray}
h(r) \simeq C_{\rm s} \sqrt{ r^3 \over G M}. 
\end{eqnarray}
In a disk where the Keplerian rotation velocity is significantly supersonic, it follows naturally that $h \ll r$.

\subsection{Growth of a purely hydrodynamic disk}
The Keplerian rotation defines a RSD. 
The rotational velocity is increased by the AM conservation during the infall, and the disk centrifugal radius grows in the absence of MB. 
Before considering the MB, We first demonstrate that a purely hydrodynamic disk grows rapidly in size. 
In the absence of MB, the accreted AM leads to the growth of disk size. We can write the equation of AM conservation at the edge of the disk
\begin{equation}
{d \rho u_\phi r \over dt} = -\rho_{\rm e} u_{{\rm e},\phi} u_r, 
\end{equation}
where $u_{{\rm e},\phi}$ is the rotation velocity in the envelope. The right-hand side of the equation describes the flux of AM from the accreting envelope into the disk edge region, while the loss through accretion inside the disk is comparably negligible. The left-hand side reflects the disk evolution due to AM accretion, with the density and rotational velocity specified by Equations (\ref{eq:rho}) and (\ref{eq:v_kep}).
The equation can be re-written as
\begin{equation}
{d\over dt} \left(\delta_\rho \delta_\phi \delta_r^2{M \over 2\pi r^3} \sqrt{GMr} \right) = -\sqrt{\beta_{\rm e}} \delta_\rho \delta_r {C_{\rm s}^2 \over 2\pi Gr^2} {GM \over r},
\end{equation}
where $\beta_{\rm e}$ is the ratio between rotational and gravitational energy in the collapsing envelope just outside the disk.
The central mass of the SIS grows at constant accretion rate and satisfies $M = 2\delta_\rho C_{\rm s}^3 t/G$ \citep{Shu1977}.
Solving the equation then yields
\begin{eqnarray}
r_{\rm hydro} &=&  {\beta_{\rm e} GM \over 4 \delta_\rho^2 \delta_\phi^2\delta_r^2 C_{\rm s}^2} \\
&=& 11.1 ~ {\rm AU} \left[{1 \over \delta_\rho^2\delta_\phi^2\delta_r^2}\right] \left[{\beta_{\rm e} \over 0.02}\right] \left[{M \over 0.1 M_\odot}\right]. \nonumber 
\end{eqnarray}
The canonical radius that we obtained appears somehow small, while it should be taken into account that the $\beta_{\rm e}$ value applied to the rotation velocity right before the envelope-disk boundary is an underestimation. 
The correction should be made for AM conservation during the infall such that
\begin{equation}
\beta_{\rm e} = \beta {r_0(M) \over r_{\rm hydro}} = \beta {GM \over 2 \delta_\rho C_{\rm s}^2 r_{\rm hydro}},
\end{equation}
where 
\begin{equation}\label{eq:r0}
r_0(M) = {GM \over 2 \delta_\rho C_{\rm s}^2}
\end{equation}
is the original radius which contains the envelope of mass $M$ and $\beta$ quantifies the amount of rotation before the collapse.
We obtain
\begin{eqnarray}\label{eq:r_hydro}
r_{\rm hydro} &=&  {\beta^{1\over 2} GM \over \sqrt{8} \delta_\rho^{3\over 2} \delta_\phi \delta_r C_{\rm s}^2} \\
&=& 111 ~ {\rm AU} \left[{1 \over \delta_\rho^{3\over 2}\delta_\phi \delta_r}\right] \left[{\beta \over 0.02}\right]^{1\over 2} \left[{M \over 0.1 M_\odot}\right]. \nonumber 
\end{eqnarray}
\citet{Hennebelle2016} found $r\propto \beta M^{1/3}\rho_0^{-2/3}$, while the same dependence on mass can be obtained by replacing the core initial density $\rho_0$ with Equation (\ref{eq:r0}). 

The dependence on $\beta$ differs in the two formula. However, the value of $\beta$ has been estimated fairly constant at $\approx 0.02$ for a wide range of prestellar cores \citep{Goodman1993, Belloche2013, Gaudel2020}, its impact should therefore be limited. Generally speaking, the size of a disk forming inside a collapsing core grows linearly with its mass, or with time, for constant $\beta$ in the absence of MB.

\subsection{Timescales and equilibria under non-ideal MHD}

\subsubsection{Equilibrium of the rotation velocity}
In contrast to the pure hydrodynamic case, a stationary state can be reached if the advected AM from the infalling envelope is canceled out by the MB.  
The stationarity here is defined on timescales much smaller than the mass growth timescale of the star-disk system, i.e., the disk rotation time, for example.
Neglecting the vertical velocity and radial magnetic field at the vicinity of the disk edge, we derive the timescale of advection\footnote{This is different from the rotation timescale used by \citet{Hennebelle2016}. However, the modified formulation would result in little difference since both the infall velocity and the rotation velocity are a fraction of the Keplerian velocity near the disk edge \citep[e.g.,][]{Marchand2020,Lee2021}. }
\begin{equation}
\tau_{\rm adv} \simeq {r \over u_r} 
\end{equation}
and the timescale of MB
\begin{equation}
\tau_{\rm br} \simeq {\rho u_\phi h \over B_z B_\phi}.
\end{equation}
Inside the envelope, the ratio between the two timescales, $\tau_{\rm adv}/ \tau_{\rm br} \propto r B_zB_\phi /(h\rho u_ru_\phi)$, increases toward the center. 
The purely hydrodynamic case corresponds to a ratio of 0.
An equilibrium, 
\begin{equation}\label{eq:tau_uphi}
\tau_{\rm adv} \simeq \tau_{\rm br},
\end{equation}
is reached when the MB counteracts the AM accretion and successfully suppresses the rapid growth of the disk size. 
The interior of the disk should be self-regulated such that this equality is roughly maintained throughout the disk evolution.

\subsubsection{Equilibrium of the azimuthal magnetic field}\label{st:tau_mag}
The azimuthal magnetic field controls the MB of the rotation. 
For conciseness, we combine the ambipolar and Ohmic diffusion into one coefficient $\eta_{\rm D} = \eta_{\rm O}+\eta_{\rm A}$ since their effects on the magnetic field dynamics are similar.

There are several mechanisms by which the azimuthal field strength could be modified: induction, advection, and diffusion. 
The equilibrium is found by equating the relevant timescales in the relevant regime.

The Hall drift velocity $\vb*{u}_{\rm H}$ describes how charged particles drift with respect to the neutrals, and thus moving the field lines with respect to the mass. This effect is sensitive to the field line configuration and can either increase or decrease the field strength.
The ambipolar or Ohmic diffusion velocity, $\vb*{u}_{\rm D}$, also moves the charged particles with respect to the neutrals, bringing the field lines with them, while this effect always smooths the magnetic field.
(Readers are invited to consult the expanded forms of the MHD equations in Appendix \ref{ap:NIMHD}.) 

Induction generates $B_\phi$ from $B_z$ through vertically differential rotation: 
\begin{equation}
\left.{\partial B_\phi \over \partial t}\right|_{\rm induction} = {\partial u_{({\rm H,})\phi} B_z \over \partial z}
\end{equation}
This induction can result from the differential rotation $u_\phi$ or the differential Hall drift $u_{{\rm H,}\phi}$
acting respectively on
the timescale of Faraday induction 
\begin{equation}
\tau_{\rm far} \simeq {B_\phi h \over B_z u_\phi},
\end{equation}
and the timescale of azimuthal Hall induction
\begin{equation}
\tau_{\rm Hall, ind} \simeq {h^2 B B_\phi \over |\eta_{\rm H}| B_z B_r}.
\end{equation}
$B_\phi$ can either increase or decrease depending on the relative orientation between $B_\phi$ and $u_\phi$. 
In the disk geometry, the induction from $B_r$ is less important and is not discussed here.

Advection of the field lines along with the flow $u_r$, or the Hall  drift $u_{{\rm H,}r}$, also changes $B_\phi$:
\begin{equation}
\left.{\partial B_\phi \over \partial t}\right|_{\rm advection} = - {\partial u_{({\rm H,})r} B_\phi \over \partial r}
\end{equation}
This yields the timescale of hydrodynamical advection, which is the same as that of the AM, 
and that of radial Hall drift
\begin{equation}
\tau_{\rm Hall, drift} \simeq { h r B \over  |\eta_{\rm H}| B_\phi},
\end{equation}
on which timescale $B_z$ is also advected.
Note that it is the differential advection, i.e., the difference between the amount entering and leaving a local volume, that varies the strength of $B_\phi$. 
The vertical velocity is less important in the vicinity of the disk and thus vertical advection is not discussed.

Finally, the ion-neutral friction and the Ohmic dissipation decrease the strength of $B_\phi$ through diffusion.
The magnetic tension straightens the field lines and behaves somehow like an inverse effect of induction: 
\begin{equation}
\left.{\partial B_\phi \over \partial t}\right|_{{\rm tensor},\phi} = {\partial u_{{\rm D},\phi} B_z \over \partial z} = -{\partial  \over \partial z} \left(\eta_{\rm D}{B_z^2 \over B^2}  {\partial B_\phi \over \partial z}\right)
\end{equation}
The magnetic tension terms straighten the field lines by azimuthal diffusion
\begin{equation}
\tau_{{\rm Diff,T},\phi} \simeq {h^2 B^2 \over \eta_{\rm D} B_z^2},
\end{equation}
The tension generated by the poloidal field not only decreases $B_r$ but also transports the $B_\phi$ component along with the diffusion:
\begin{equation}
\left.{\partial B_\phi \over \partial t}\right|_{{\rm tensor},r} = {\partial u_{{\rm D},\phi} B_r \over \partial r} = -{\partial  \over \partial r} \left(\eta_{\rm D}{B_zB_r \over B^2}  {\partial B_\phi \over \partial z}\right),
\end{equation}
which yield the timescale for radial diffusion (same timescale for $B_z$)
\begin{equation}
\tau_{{\rm Diff,T},r} \simeq {hr B^2 \over \eta_{\rm D} B_z B_r}.
\end{equation}

The magnetic pressure terms leads to the dilution of the field lines down the pressure gradient. 
The vertical diffusion relaxes the winded toroidal field to a larger vertical extent:
\begin{eqnarray}
\left.{\partial B_\phi \over \partial t}\right|_{{\rm pressure},z} &=& -{\partial u_{{\rm D},z} B_\phi \over \partial z} \\
&=& {\partial  \over \partial z} \left[{\eta_{\rm D}\over B^2}\left(B_\phi^2  {\partial B_\phi \over \partial z}+B_\phi B_r  {\partial B_r \over \partial z}\right)\right], \nonumber
\end{eqnarray}
which yields the timescale vertical diffusion (same for $B_r$)
\begin{equation}
\tau_{{\rm Diff, P},z} \simeq {h^2 B^2 \over \eta_{\rm D}  (B_\phi^2 + B_r^2)},
\end{equation}
The diffusion also happens in the radial direction:
\begin{eqnarray}
\left.{\partial B_\phi \over \partial t}\right|_{{\rm pressure},r} &=& -{\partial u_{{\rm D},r} B_\phi \over \partial r} \\
&=& {\partial  \over \partial r} \left[{\eta_{\rm D}\over B^2}\left({B_\phi^2 \over r} {\partial r B_\phi \over \partial z}+B_\phi B_z  {\partial B_z \over \partial r}\right)\right],\nonumber
\end{eqnarray}
yielding the timescale for radial diffusion (same for $B_z$)
\begin{equation}
\tau_{{\rm Diff, P},r} \simeq {r^2 B^2 \over \eta_{\rm D} (B_\phi^2 + B_z^2)}.
\end{equation}
The radial pressure diffusion should be of less importance, except for the envelope, where the geometry is not yet flattened.

\subsection{Dependencies of the Equilibrium Disk Radius}\label{sec:cases}
In the cases where the Hall effect is dynamically important, one has to carefully consider the sign of the Hall resistivity coefficient, $\eta_{\rm H}$, to see whether an equilibrium could be reached. 
The sign change of $\eta_{\rm H}$ near density $n \sim 10^{13}~ {\rm cm}^{-3}$ 
is mostly due to the transition of dominance by positively charged ions at low density and electrons at high density \citep{Marchand2016,Zhao2020a}. 

Equation (\ref{eq:tau_uphi})
is always satisfied inside the disk since the rotation velocity cannot increase beyond the Keplerian value.
We discuss the different regimes of $\lambda$ and $\beta$ and find the relevant timescales below (with initial field in $+z$ direction): strong field where $B_z$ dominates inside the disk versus weak field where the field lines are winded by the rotation and $B_\phi$ dominates, low rotation where the spin direction is set by the Hall drift irrespective of the initial rotation versus high rotation where the disk preserves core initial rotation direction. Exact values separating those regimes are discussed in Section \ref{sec:regimes}. The factors $\delta_\rho$, $\delta_\phi$, and $\delta_r$ are of order unity and will be omitted in the following text for conciseness, given that the dependencies are at most linear, while sublinear most of the time.

\subsubsection{Strong field, low rotation case}\label{sec:sB_wR}
With small $\lambda$ and small $\beta$, the collapsing envelope pinches the field lines in the radial direction: 
$$B_z > B_r > B_\phi, B\approx B_z.$$

At this density range, $\eta_{\rm H}<0$. 
The Hall drift generates an azimuthal component that will torque up the collapsing gas in $-z$ direction.
Such rotation anti-parallel to the magnetic field ($\vb{B}\cdot \vb{\Omega}<0$) will be further amplified by the collapse. 
With increasing $u_\phi$, the Faraday induction from vertical differential rotation will dominate over the Hall induction in generating $B_\phi$, and the direction of $B_\phi$ will be eventually reversed.

The azimuthal velocity in turn drives the inward radial Hall drift and lowers the local mass-to-flux ratio. Note that the flux is lowered, not increased, due to differential inward drift. 
An equilibrium is reached when
\begin{eqnarray}\label{eq:tau_RSHCS}
\tau_{\rm far} \simeq \tau_{\rm Hall, drift}.
\end{eqnarray}
Combining Equations (\ref{eq:tau_uphi}) and (\ref{eq:tau_RSHCS}), the corresponding radius is 
\begin{eqnarray}\label{eq:rO_sB_wR}
r_{\rm O} &=& \left[ (2\pi)^{-4}G^{1}C_{\rm s}^{4} |\eta_{\rm H}|^{2} M^{5}B_z^{-8}\right]^{1 \over 15} \\
&=&  13.0 ~ {\rm AU}~ 
\left[ {|\eta_{\rm H}| \over 10^{19} {\rm cm}^2~{\rm s}^{-1}}\right]^{2 \over 15} \left[{M \over 0.1 M_\odot}\right]^{1 \over 3} \left[{B_z \over 0.1 {\rm G}}\right]^{-{8 \over 15}}. \nonumber
\end{eqnarray}
This disk radius is $\lesssim 20$ AU for sub-Solar mass, with sub-linear dependence on the mass and vertical field strength. The dependence on the Hall resistivity is very weak with an exponent of $2/15$.
The vertical Hall drift is toward the midplane and thus this structure will eventually flatten.
This is referred to by \citet{Zhao2020a} as the rotationally supported Hall current sheet (RSHCS).

The magnetic flux in the outer envelope can be lost through radial diffusion (mostly ambipolar at low density) on $\tau_{{\rm Diff, T},r}$ (see Appendix \ref{ap:Br} for the discussion of radial diffusion and strength of $B_r$), and $B_\phi$ is transported along. However, this loss is too weak to compensate the Faraday induction. Indeed, the solution $r_{\rm O}$ satisfies the presumption 
\begin{eqnarray}\label{eq:Hall_vs_Diff}
\tau_{\rm Hall, drift} < \tau_{{\rm Diff, T},r}.
\end{eqnarray}
Detailed consistency check for all simplifying presumptions are presented in Appendix \ref{ap:consistency}.

Further inwards, $|\eta_{\rm H}|$ decreases with increasing density and approaches zero. 
At such radius, 
the equilibrium is established instead by
\begin{eqnarray}
\tau_{\rm far} \simeq \tau_{{\rm Diff, T},\phi},
\end{eqnarray}
that is, the drag of field lines in the azimuthal direction is balanced by the diffusion in opposite direction. 
This gives the radius
\begin{eqnarray}\label{eq:rI_sB_wR}
r_{\rm I} &=& \left[ (2\pi)^{-2}G^{1}C_{\rm s}^{0} \eta_{\rm D}^{2} M^{3}B_z^{-4} \right]^{1\over9}\\
&=&  19.2 ~ {\rm AU}  
\left[{\eta_{\rm D} \over 10^{19} {\rm cm}^2~{\rm s}^{-1}}\right]^{2\over9} \left[{M \over 0.1 M_\odot}\right]^{1\over3}\left[ {B_z \over 0.1 {\rm G}}\right]^{-{4\over9}}. \nonumber
\end{eqnarray}
This corresponds to the RSD \citep[][]{Zhao2020a} or the para-disk \citep{Tsukamoto2015a}. 
This formula is exactly the same as that derived by \citet{Hennebelle2016}, while at such density, the relevant diffusion is likely the Ohmic diffusion.
This radius has the same dependence on $M$ as that of $r_{\rm O}$, while its dependence on $B_z$ is even shallower. The results depends only weakly on the diffusion resistivity. 

By construction, 
\begin{eqnarray}\label{eq:rO_rI}
r_{\rm O}>r_{\rm I}
\end{eqnarray}
and $B_z(r)$ is a decreasing function.
Equation (\ref{eq:rO_rI}) is not satisfied for all combinations of mass and vertical field. 
Dividing Equation (\ref{eq:rO_sB_wR}) by equation (\ref{eq:rI_sB_wR}) yields
\begin{eqnarray}
{r_{\rm O} \over r_{\rm I} }
&=&\left[ (2\pi)^{-2}G^{-2}C_{\rm s}^{12} {|\eta_{\rm H}|(r_{\rm O})^{6} \over \eta_{\rm D}(r_{\rm I})^{10}} {M(r_{\rm O})^{15} \over M(r_{\rm I})^{15}} {B_z(r_{\rm I})^{20} \over B_z(r_{\rm O})^{24}} 
\right]^{1\over45}\\
&=& 0.68 ~ 
\left[{|\eta_{\rm H}(r_{\rm O})| \over 10^{19} {\rm cm}^2~{\rm s}^{-1}}\right]^{2 \over 15}\left[{\eta_{\rm D}(r_{\rm I}) \over 10^{19} {\rm cm}^2~{\rm s}^{-1}}\right]^{-{2\over9}} \times \nonumber \\&&
\left[{M(r_{\rm O}) \over M(r_{\rm I})}\right]^{1\over 3} \left[{B_z(r_{\rm I}) \over B_z(r_{\rm O})}\right]^{4\over 9} 
\left[{B_z(r_{\rm O}) \over 0.1 {\rm G}}\right]^{-{4 \over 45}}. \nonumber
\end{eqnarray}

The magnetic flux accumulates with the inward drifting of the field lines, and the two radii will approach each other. 
Inside $r_{\rm I}$, the magnetic flux is diffused outward and accumulates near $r_{\rm I}$, naturally creates a contrast in $B_z$ values at the two radii, which is seen in simulations \citep[e.g. Fig. 11 of][]{Zhao2020a}. 
Even without considering the fact that $M(r_{\rm O}) > M(r_{\rm I})$, a ratio $B_z(r_{\rm I})/B_z(r_{\rm O})\gtrsim 2.4$ is easily guaranteed \citep[see Fig. 5 in][]{Zhao2020a} and is sufficient to satisfy the condition (\ref{eq:rO_rI}). 
The increased field strength allows $|\eta_{\rm H}|$ to maintain a non-negligible value as the density increases, while at the same time, the piling up of magnetic flux around $r_{\rm I}$ decreases the efficiency of radial Hall drift and $r_{\rm O}$ shrinks in consequence.
The condition $\tau_{\rm Hall, drift} < \tau_{{\rm Diff, T},\phi}$ for $r_{\rm O}$ will eventually be violated. The RSHCS will disappear and only the RSD is long-lived, with the disk radius roughly corresponding to the sign inversion of $\eta_{\rm H}$. More precisely, a small discrepancy between $r_{\rm O}$ and $r_{\rm I}$ always persists and reflects the maximum ratio of $B_z$ values that can be sustained under non-ideal effects near the disk edge.

The Ohmic dissipation starts to be important when $\eta_{\rm H}$ approaches zero rapidly and is about to change sign. 
The density at which the sign of $\eta_{\rm H}$ changes follows roughly the relation 
\begin{eqnarray}\label{eq:rhoB}
n_{\rm t}(B) = {\rho_{\rm t}(B) \over \mu m_{\rm p}} = 8 \times 10^{13}~ {\rm cm}^{-3} {B \over 1~ {\rm G}} = k B,
\end{eqnarray}
where we use $\mu=2.3$ for the mean molecular weight and $m_{\rm p}$ is the proton mass. This expression has been determined using the chemical abundance table computed by \citet{Marchand2016}, and we provide an analytic justification in Appendix \ref{ap:rho_crit}.
Therefore, we can solve the equation
\begin{eqnarray}\label{eq:rd_rhot_rhoB}
\rho(r) = \rho_{\rm t}(B(r)),
\end{eqnarray}
to approximately find the equilibrium radius. 
In the regime dominated by the vertical field, the radius becomes
\begin{eqnarray}\label{eq:rt_sB}
r_{\rm t} &=&  \left({  M\over 2\pi k\mu m_{\rm p} B_z }\right)^{1\over3} \\
&=&  6.75 ~ {\rm AU} 
\left[{k \over 8~ 10^{13} {\rm cm}^{-3}{\rm G}^{-1}}\right]^{-{1\over3}}  \left[{M \over 0.1 M_\odot}\right]^{1\over3} \left[{B_z \over 0.1 {\rm G}}\right]^{-{1\over3}}. \nonumber 
\end{eqnarray}
Equation (\ref{eq:rt_sB}) is very similar to Equation (\ref{eq:rI_sB_wR}), with only a slight difference by $1/9$ in the $B_z$ exponent, while the numerical value is smaller. 
In fact, the AD/Ohmic resistivity can dominate over the Hall resistivity well before $\eta_{\rm H}$ actually reaches zero \citep[see Fig.5 in][]{Marchand2016}. Lowering $n_{\rm t}$ by a factor $10-100$ easily removes this discrepancy.

\subsubsection{Strong field, high rotation case}
With small $\lambda$ and large $\beta$, the field lines are more pinched azimuthally than in the radial direction: 
$$B_z > B_\phi > B_r, B\approx B_z.$$

The anti-parallel case ($\vb{B}\cdot \vb{\Omega}<0$) is similar to the low rotation case, while the field inward drifting region is more extended.
On the other hand, in the parallel case, the outward Hall drift increases the azimuthal field strength (again, through differential drift, possibly faster than the Faraday induction).
Solving the equality
\begin{eqnarray}\label{eq:tau_Rortho}
\tau_{\rm Hall, drift} \simeq \tau_{{\rm Diff, T},\phi}
\end{eqnarray}
yields the radius
\begin{eqnarray}\label{eq:rOrtho_sB}
r_{\rm ortho} &=& \left[ (2\pi)^{-1} C_{\rm s}^{2} |\eta_{\rm H}|  \eta_{\rm D}^{-1}  M B_z ^{-2}\right]^{1\over 3}\\
&=&7.22 ~ {\rm AU}~  \left[{|\eta_{\rm H}| \over \eta_{\rm D}}\right]^{1\over3}\left[ {M \over 0.1 M_\odot}\right]^{1\over3} \left[{B_z \over 0.1 {\rm G}}\right]^{-{2\over3}}. \nonumber
\end{eqnarray}
This is referred to as the ortho-disk by \citet{Tsukamoto2015a}, which is smaller than the para-disk under otherwise similar conditions.
Recalling the previous discussions of the anti-parallel configuration, the Hall drift and the azimuthal diffusion are used both to balance against the Faraday induction, at $r_{\rm O}$ and $r_{\rm I}$, respectively. Only the parallel configuration allows these two mechanisms to cancel out each other.

This solution is valid only if the density at the edge of the disk is lower than the transition density, $n_{\rm t}$ (Equation (\ref{eq:rhoB})), and $\eta_{\rm H} < 0$.
As the evolution proceeds, the flux piles up at the disk edge, increasing $n_{\rm t}$ locally, and the region of positive $\eta_{\rm H}$ shrinks in consequence. 
The magnetic field amplified by the Hall inward drift in the inner disk ($\eta_{\rm H} > 0$) strongly brakes the rotation, and the disk will become sub-Keplerian. 
The two effects act in synergy to rapidly redistribute the AM and decrease the ortho-disk size, which will disappear eventually.
After the disappearance of the ortho-disk, the gas rotation is accelerated in the opposite direction and a para-disk will form as in the low rotation case. 
This behavior is observed by \citet{Zhao2020a}, which means that their initial rotation is strong enough (for how strong see Section \ref{sec:regimes}).

\subsubsection{Weak field, low rotation case}
With large $\lambda$ and small $\beta$, 
the radial collapse in the envelope strongly pinches the field lines: 
$$ B_\phi > B_r >B_z , B\approx B_\phi. $$
The radial field component in turn generates an azimuthal field that torques up the gas in the opposite direction of the vertical field, and eventually a para-disk forms.
Detailed justification that $B_\phi > B_r$ is discussed in Appendix \ref{ap:Br}.
We solve the same equation (\ref{eq:tau_RSHCS}) for the RSHCS in this regime,
which gives
\begin{eqnarray}\label{eq:rO_wB_wR}
r_{\rm O} &=& \left[  (2\pi)^{-1}G^{0}C_{\rm s}^{1} |\eta_{\rm H}|^{1}  M^{1}B_z^{-2} \right]^{1\over4} \\
&=&  10.6 ~ {\rm AU}~ 
\left[{|\eta_{\rm H}| \over 10^{19} {\rm cm}^2~{\rm s}^{-1}}\right]^{1\over4} \left[{M \over 0.1 M_\odot}\right]^{1\over4} \left[{B_z \over 0.1 {\rm G}}\right]^{-{1\over2}}. \nonumber
\end{eqnarray}
In the inner region where $\eta_{\rm H}$ vanishes, the dominant loss term is the vertical diffusion, which gives 
\begin{eqnarray}
\tau_{\rm far} \simeq \tau_{{\rm Diff, P},z}.
\end{eqnarray}
This yields exactly the same solution for the RSD as in the strong field case. 

Likewise, Equation (\ref{eq:rO_rI}) should be satisfied by construction. This allows only certain combinations of $M(r_{\rm O})$, $M(r_{\rm I})$, $B_z(r_{\rm O})$, and $B_z(r_{\rm I})$. 
Combining Equations (\ref{eq:rO_wB_wR}) and (\ref{eq:rI_sB_wR}) gives
\begin{eqnarray}
{r_{\rm O} \over r_{\rm I}}
&=&\left[ (2\pi)^{-1}G^{-4}C_{\rm s}^{9} { |\eta_{\rm H}|(r_{\rm O})^{9} \over \eta_{\rm D}(r_{\rm I})^{8}} {M(r_{\rm O})^{9} \over M(r_{\rm I})^{12}}  {B_z(r_{\rm I})^{16} \over B_z(r_{\rm O})^{18}}
\right]^{1\over36}\\
&=& 0.55 ~
 \left[{|\eta_{\rm H}(r_{\rm O})| \over 10^{19} {\rm cm}^2~{\rm s}^{-1}}\right]^{1 \over 4}\left[{\eta_{\rm D}(r_{\rm I}) \over 10^{19} {\rm cm}^2~{\rm s}^{-1}}\right]^{-{2\over9}} \times \nonumber \\&& 
\left[{M(r_{\rm O}) \over M(r_{\rm I})}\right]^{1\over 3} \left[{B_z(r_{\rm I}) \over B_z(r_{\rm O})}\right]^{4\over 9} 
\left[{M(r_{\rm O}) \over 0.1 M_\odot}\right]^{-{1\over 12}}\left[{B_z(r_{\rm O}) \over 0.1 {\rm G}}\right]^{-{1 \over 18}}. \nonumber
\end{eqnarray}
Similarly to the strong field case, the outward diffusion and inward Hall drift will always maintain a contrast in $B_z$ values even when $r_{\rm O}$ and $r_{\rm I}$ becomes very close to each other. 
A ratio of $B_z(r_{\rm I})/ B_z(r_{\rm O}) \gtrsim 3.8$ guarantees the presumed Equation (\ref{eq:rO_rI}).
The RSHCS will disappear as described in the strong field case when the accumulated magnetic flux diminishes the Hall effect, and only the RSD will survive.

Solving Equation (\ref{eq:rd_rhot_rhoB}) in this regime yields 
the radius at which $\eta_{\rm H}$ changes sign:
\begin{eqnarray}
r_{\rm t} &=&   C_{\rm s}^2 k^2 G M B_z^{-2} \\
&=&  3345 ~ {\rm AU}~  
\left[{k \over 8~ 10^{13} {\rm cm}^{-3}{\rm G}^{-1}}\right]^{2}\left[ {M \over 0.1 M_\odot}\right]^{1} \left[{B_z \over 0.1 {\rm G}}\right]^{-2}. \nonumber 
\end{eqnarray}
This expression has super-linear dependence on the parameters, 
 while $M$ and $B_z$ both increase during the disk growth. 
It is thus not straightforward to interpret the size evolution.
The alternative expression obtained by inserting Equation (\ref{eq:lambda}) is therefore more appropriate:
\begin{eqnarray}\label{eq:rd_ortho_approx}
r_{\rm t} = 
18.5 ~ {\rm AU} 
\left[{k \over 8~ 10^{13} {\rm cm}^{3}{\rm G}^{-1}}\right]^{2} \left[{M \over 0.1 M_\odot}\right]^{1\over3}\tilde{\lambda}^{-{2\over3}},
\end{eqnarray}
where $\tilde{\lambda}$ is calculated with the local $B_z$ at the disk edge instead of the global disk flux. This should be lower than the disk global $\lambda$ since the field lines accumulate at the disk edge due to outward diffusion and inward Hall drift \citep[see e.g. Fig.11 in][]{Zhao2020a}. 
\citet{Masson2016} found values of $\lambda$ several tens to a hundred in the presence of ambipolar diffusion. 
This expression then gives very reasonable disk size for $\tilde{\lambda}$ of a few, up to a few tens for significantly supercritical cores.

\subsubsection{Weak field, high rotation case}
With large $\lambda$ and $\beta$, 
the anti-parallel case forms again a para-disk very similar to that in the low rotation case. 
As for the ortho case, we solve
\begin{eqnarray}
\tau_{\rm Hall, drift} \simeq \tau_{{\rm Diff, P},z}, 
\end{eqnarray}
which yields 
\begin{eqnarray}\label{eq:rOrtho_wB}
r_{\rm ortho} &=& G C_{\rm s}^{-2} \eta_{\rm D}^2 |\eta_{\rm H}|^{-2}  M \\
&=&  2223 ~ {\rm AU} \left[{\eta_{\rm D} \over |\eta_{\rm H}| }\right]^{2} \left[{M \over 0.1 M_\odot}\right]. \nonumber
\end{eqnarray}
In the density range where the equilibrium is reached, $\eta_{\rm D}/|\eta_{\rm H}|$ increases with increasing density, and thus decreasing radius. 
A ratio around 1/10 yields roughly a disk size $\sim 20$ AU, while the exact size depends on the detailed chemical balancing.

Like in the strong field case, local magnetic flux at the disk edge increases under accumulation and $n_{\rm t}$ increases in consequence. This will increase $|\eta_{\rm H}|$, therefore reducing the ortho-disk size.  After the disappearance of the ortho-disk, a counter-rotating disk will eventually form as in the weak rotation case.

It is possible to further estimate this radius. 
From Equation (\ref{eq:etaH_etaD}), we can estimate the ratio between the two resistivities at density $n<n_{\rm t}$:
\begin{eqnarray}\label{eq:etaD_etaH}
{\eta_{\rm A} + \eta_{\rm O} \over \eta_{\rm H}} \approx {n \over n_{\rm t}}  \left(n_{\rm M} \over n_{\rm e}\right)^{1 \over 2}
\end{eqnarray}

Inserting into Equation (\ref{eq:rOrtho_wB}) yields
\begin{eqnarray}\label{eq:BzOrtho_wB}
B_{z, {\rm ortho}} &=&  C_{\rm s}^2 k \mu m_{\rm p} \left(n_{\rm e} \over n_{\rm M}\right)^{1 \over 2} \\
&=& 0.011~{\rm G} ~  \left[{k \over 8~ 10^{13} {\rm cm}^{3}{\rm G}^{-1}}\right], \nonumber
\end{eqnarray}
where number density ratio between ions and electrons $n_{\rm M}/n_{\rm e} \approx 107$ is applied \citep{Marchand2021}. 
Since the Hall effect drifts the magnetic field outward, the equilibrium radius shrinks as the local value of $B_z$ decreases.
Inserting Equation (\ref{eq:lambda}), the radius can be written as
\begin{eqnarray}\label{eq:rOrtho_wB_2}
r_{\rm ortho} &=& \sqrt{2} ( {k \mu m_{\rm p}})^{-{1\over2}} G^{1\over4}C_{\rm s}^{-1}\left(n_{\rm M} \over n_{\rm e}\right)^{1 \over 4} M^{1\over2} \tilde{\lambda}^{-{1\over2}} \\
&=& 197~{\rm AU}~ \left[{k \over 8~ 10^{13} {\rm cm}^{3}{\rm G}^{-1}}\right]^{-{1\over2}} \left[{M \over 0.1 M_\odot}\right]^{1\over2}\tilde{\lambda}^{-{1\over2}}.\nonumber
\end{eqnarray}
In the case of ortho-RSD, flux piling up at the disk edge is less serious since both the diffusion and the Hall drift are outward, and thus $\tilde{\lambda}$ is relatively a good approximation of $\lambda$.
The ortho-RSD has constantly decreasing flux and mass, and thus $\lambda$ grows easily to large values. Taking $\lambda = 100$ yields a disk radius $<20$ AU.

\section{Discussions}\label{sec:discussion}

\begin{deluxetable*}{ccDccccc}
\tablecaption{Summary of the radius dependencies on disk parameters\label{tab:radius}}
\tablewidth{0pt}
\tablehead{
\colhead{Type} & \colhead{Dominant $B$ component} & \multicolumn2c{Radius} & \colhead{$M$-exponent} &\colhead{$B_z$-exponent} &\colhead{$\lambda$-exponent} & \colhead{$\eta_{\rm H}$-exponent} &\colhead{$\eta_{\rm D}$-exponent}  \\
\colhead{} & \colhead{} & \multicolumn2c{  (AU)  }  & \colhead{($0.1~M_\odot$)} & \colhead{($0.1~G$)}  & \colhead{} &\colhead{($10^{19}~{\rm cm}^2~{\rm s}^{-1}$)} & \colhead{($10^{19}~{\rm cm}^2~{\rm s}^{-1}$)}
}
\decimals
\startdata
Hydro & - & 111 & 1 & & & & \\
RSHCS & $B_z$ & 13.0  & 1/3 & -8/15 && 2/15 & \\
RSHCS & $B_\phi$ & 10.6 & 1/4 & -1/2 && 1/4 &\\
para-RSD &  $B_z/B_\phi$ & 19.2 & 1/3 & -4/9 & && 2/9\\
$r_{\rm t}$  & $B_z$ & 6.75 & 1/3 & -1/3 &\\
$r_{\rm t}(\lambda)$  & $B_z$  & \textcolor{gray}{0.0670} & 0 && 1 \\
$r_{\rm t}$  & $B_\phi$ &  \textcolor{gray}{3345} & 1 & -2 & \\
$r_{\rm t}(\lambda)$  & $B_\phi$ & 18.5 & 1/3 & & -2/3 \\
ortho-RSD & $B_z$ & 7.22 & 1/3 &  -2/3 & & 1/3 & -1/3\\
ortho-RSD & $B_\phi$ & \textcolor{gray}{2223} & 1 &   & & -2 &  2  \\
ortho-RSD & $B_z/B_\phi$ & 197 & 1/2 &   & -1/2 &  &   \\
\enddata
\tablecomments{In the cases where the initial rotation is small or anti-parallel to the magnetic field, an outer disk (RSHCS) and an inner disk (para-RSD) will form. The RSHCS solution only exists for low mass and will eventually disappear. On the other hand the para-RSD is relatively long-lived. This radius also corresponds to the radius, $r_{\rm t}$, at which $\eta_{\rm H}$ changes sign, which is expressed as functions of either $B_z$ or $
\lambda$. The ortho-RSD only forms if the initial rotation is strong and parallel to the magnetic field. This equilibrium is unstable and the ortho-RSD will eventually disappear. The dependencies on the disk parameters are weak in both strongly and weakly magnetized regimes, and the solutions are degenerate for the para-RSD, $r_{\rm I}(M,B_z)$, and the ortho-RSD, $r_{\rm ortho}(M,\lambda)$. When a proper expression with sub-linear dependencies on disk parameters is found, }a disk of radius 10-20 AU always forms in a magnetized collapsing core in the presence of NIMHD effects, modulo the weak dependencies on the uncontrolled numerical factors ($\delta_{\rho, \phi, r}$, of order unity). 
\end{deluxetable*}

We discussed four regimes of protoplanetary disk formation inside a collapsing, magnetized prestellar core. 
The disk radius was derived under different conditions of local balance. 
We derived formulae for the disk radius in a moderately magnetized collapsing core, which depends on sub-linear combinations of $M$ and $B_z$, or $M$ and $\lambda$.
We showed that regardless of a wide range of initial conditions (magnetization and rotation) of the prestellar core, 
a disk of size $\sim 10-20$ AU always forms.
If the magnetic field is too weak, an almost pure hydrodynamic disk forms and grows rapidly. 

\begin{figure*}
\setlength{\unitlength}{\textwidth}
\begin{picture}(1,0.64)
\put(0,0.32){\includegraphics[trim=200 200 150 150,clip, width=0.5\textwidth]{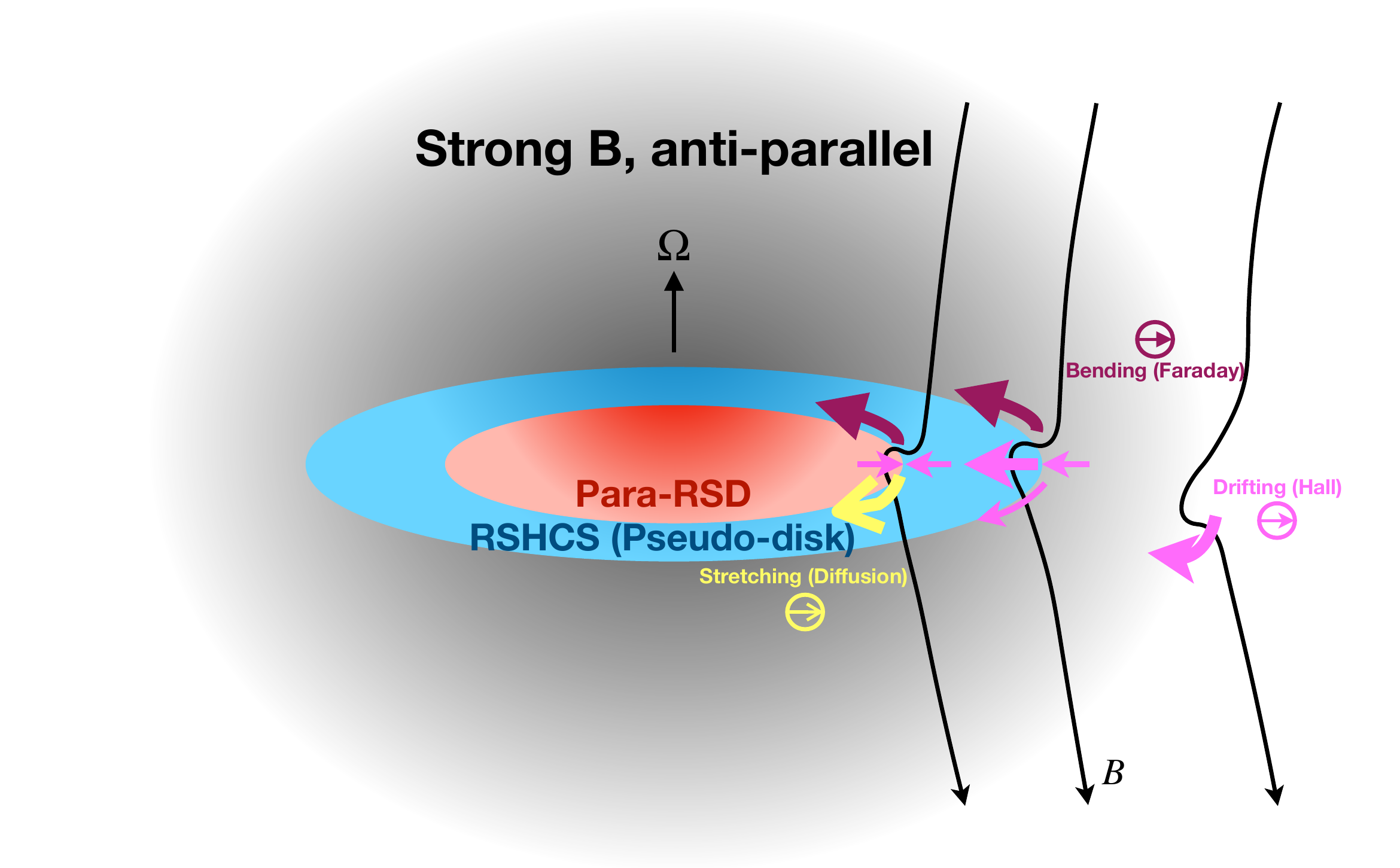}}
\put(0.5,0.32){\includegraphics[trim=200 200 150 150,clip, width=0.5\textwidth]{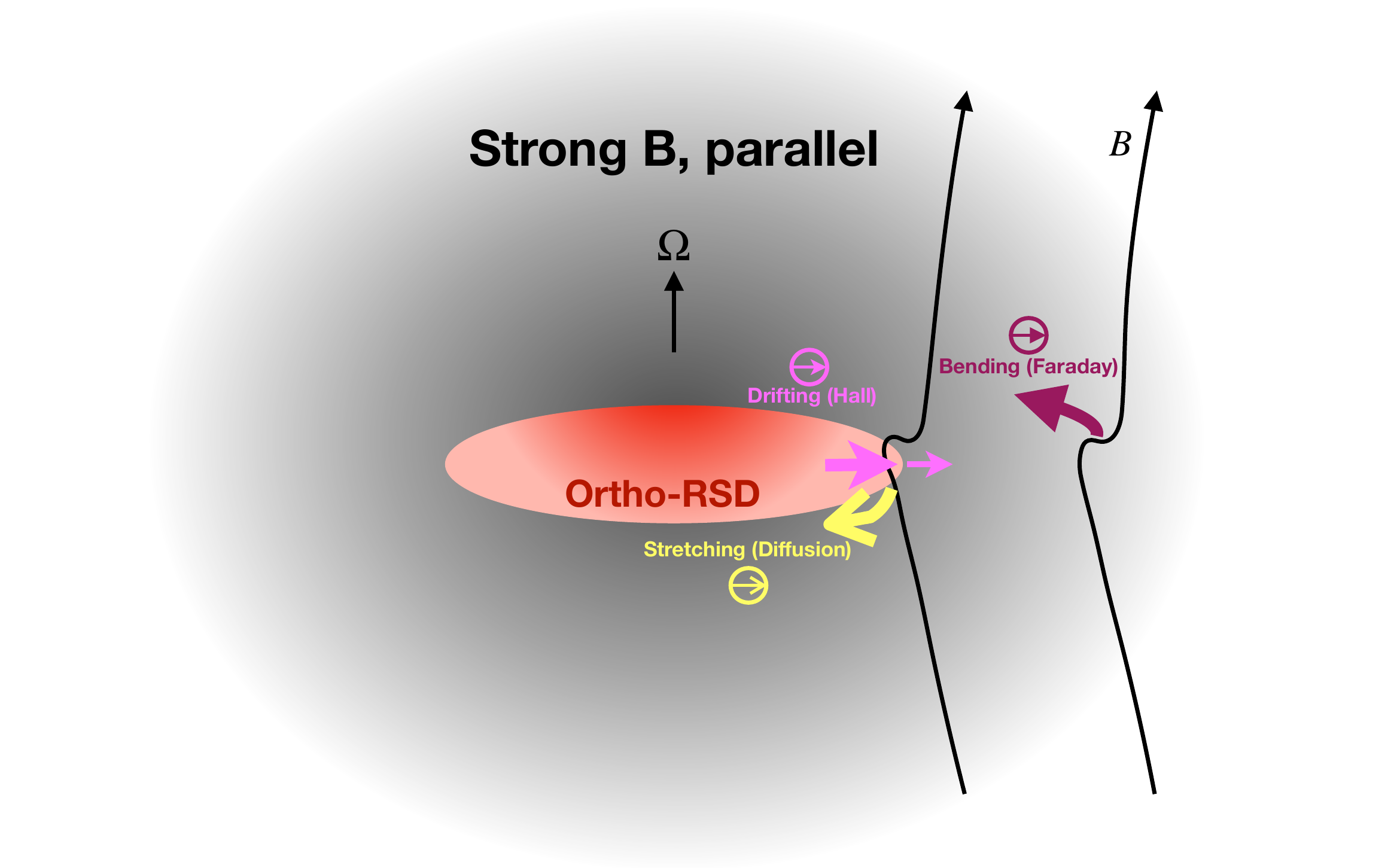}}
\put(0,0){\includegraphics[trim=200 200 150 150,clip, width=0.5\textwidth]{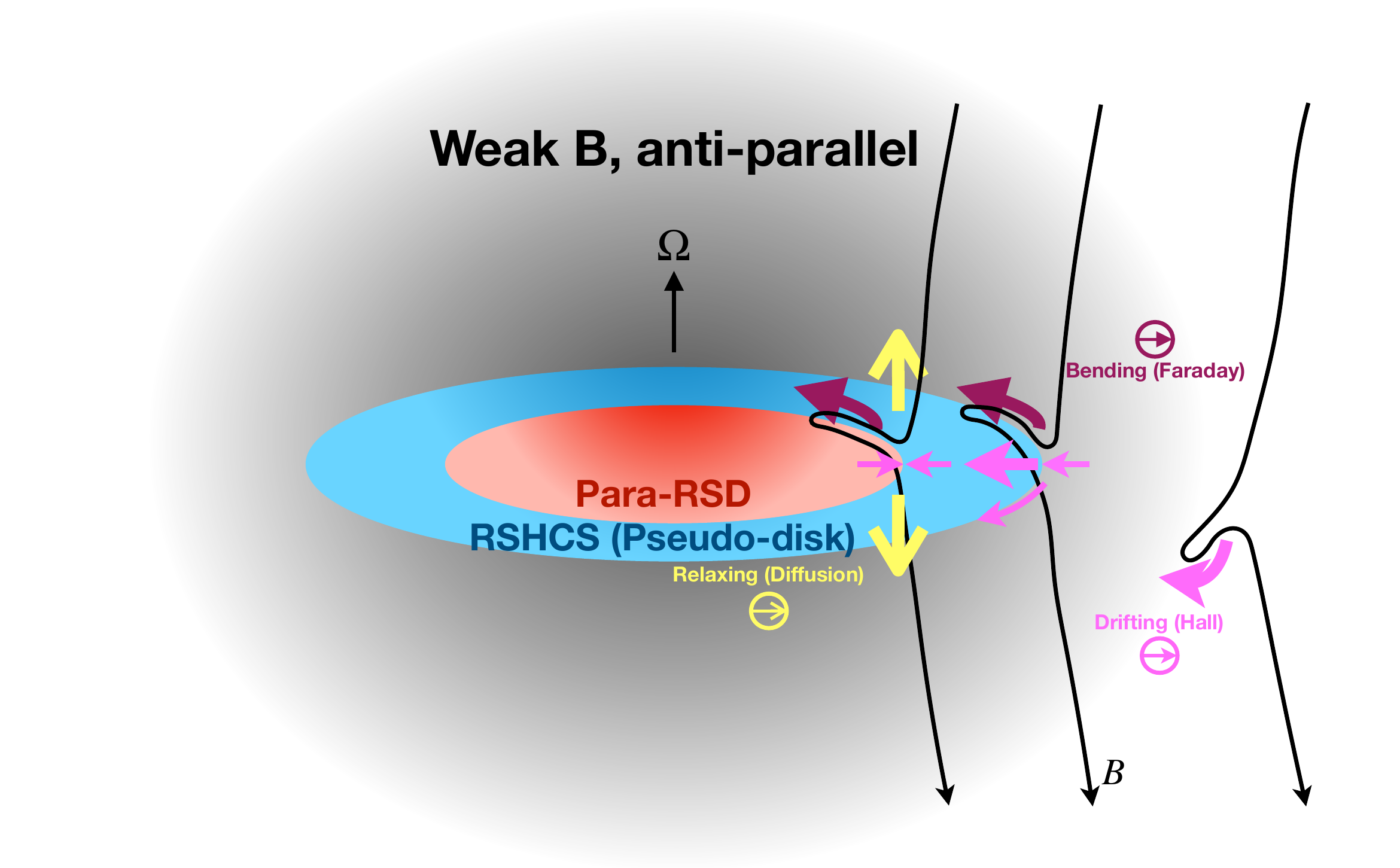}}
\put(0.5,0){\includegraphics[trim=200 200 150 150,clip, width=0.5\textwidth]{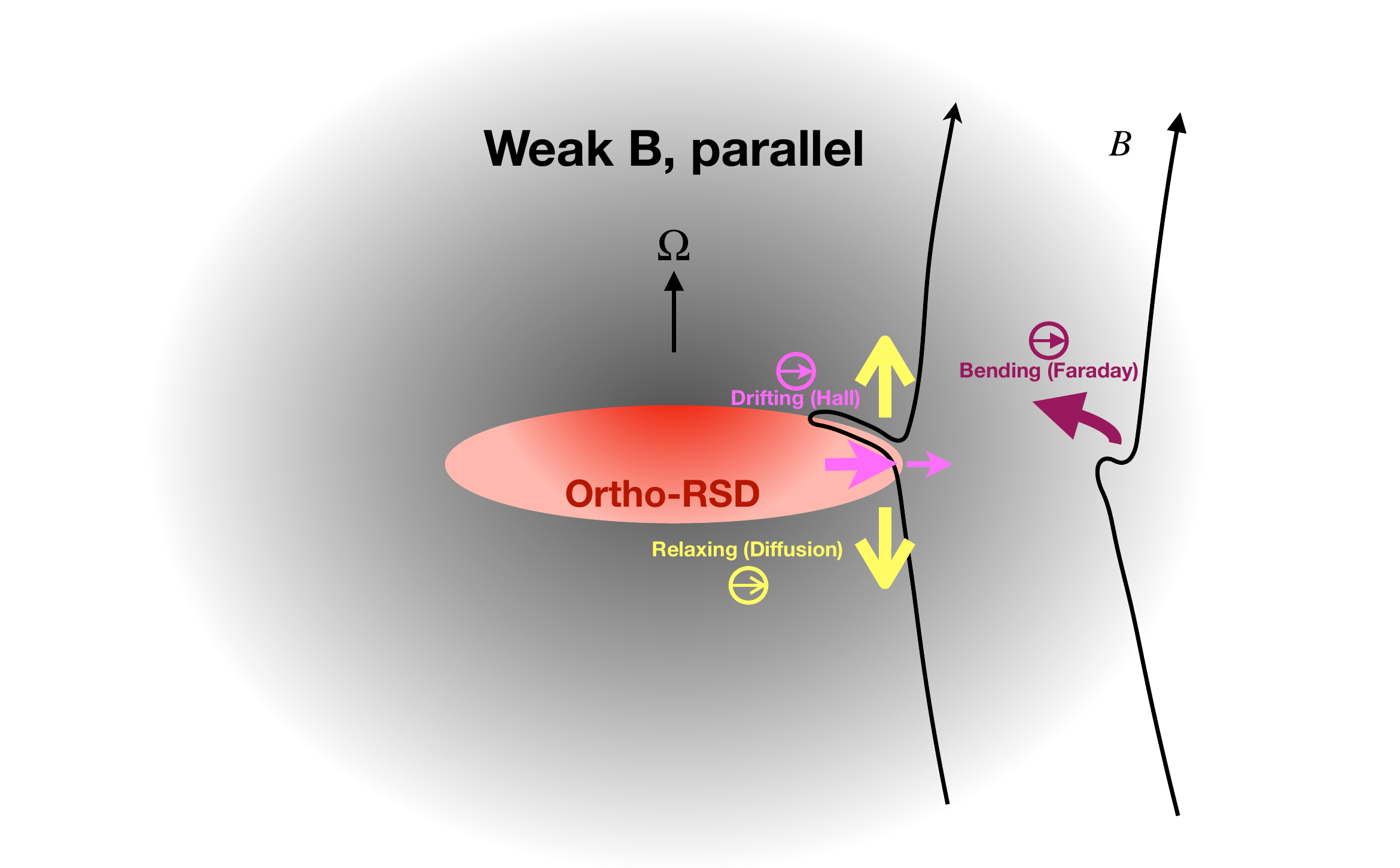}}
\put(0.1,0.32){(a)}
\put(0.6,0.32){(b)}
\put(0.1,0){(c)}
\put(0.6,0){(d)}
\end{picture}
\caption{Schematic plots for the four cases discussed. The three main mechanisms, including Faraday induction through differential rotation, Hall drift, and diffusion, are shown with arrows of different colors and types. The thick arrows are the major acting mechanisms that reach equilibrium at respective positions.\label{fig:schema}}
\end{figure*}

The results are consistent with previous works \citep[e.g.,][]{Tsukamoto2015a, Hennebelle2016, Masson2016, Wurster2016, Marchand2019, Zhao2020a, Zhao2020b}. Figure \ref{fig:schema} shows the schematic plots for the four cases, and we summarize the disk formation behavior under the non-ideal MHD effects in Table \ref{tab:radius}.

\subsection{Separation of different regimes}\label{sec:regimes}

Regimes of different $\lambda$ and $\beta$ are discussed in Section \ref{sec:cases}. 
Here, we further discuss what values separate the regimes. For clarity, we label the variables with suffixes s/w for strong/weak field cases.

\subsubsection{Para-disk}
We first discuss the long-lived para-RSD, which is a more relevant outcome related to the core initial conditions. 
Combining Equations (\ref{eq:lambda}) and (\ref{eq:rI_sB_wR}) for the para-RSD yields:
\begin{eqnarray}\label{eq:lambda_rI}
\lambda_{\rm I}  &=& 2\left[ (2 \pi)^2 G^{1} \eta_{\rm O}^{-2}  M r_{\rm I} \right]^{1\over4} \\
&=& 10.6 ~ \left[{ \eta_{\rm O}\over 10^{19} {\rm cm}^2~{\rm s}^{-1}} \right]^{-{1\over2}} \left[{M \over 0.1 M_\odot}\right]^{1\over4}\left[{r_{\rm O} \over 10 {\rm AU} }\right]^{{1\over4}}. \nonumber
\end{eqnarray}
Inserting strong vertical field presumption
\begin{eqnarray}\label{eq:cond_sB}
B_z > B_\phi
\end{eqnarray}
yields
\begin{eqnarray}\label{eq:lambda_rI_cond_sB}
\lambda_{\rm I,s}  &<& 2\left[ (2 \pi)^2 G^{1} C_{\rm s}^{-1} \eta_{\rm O}^{-1} M \right]^{1\over4} \\
&=& 14.3 ~ \left[{ \eta_{\rm O}\over 10^{19} {\rm cm}^2~{\rm s}^{-1}} \right]^{-{1\over4}} \left[{M \over 0.1 M_\odot}\right]^{1\over4}. \nonumber
\end{eqnarray}
The para-RSD in the weak $B_z$ case needs to satisfy exactly the opposite condition. 
As $\lambda$ reaches usually a few tens to a hundred inside the RSD in most of the existing simulations, the azimuthal field component likely dominates over the vertical component regardless of initial conditions.
Recall that  \citet{Zhao2020a} used $\lambda \sim 2.4$ and $4.8$ for their strong and weak field cases. 
These are however initial values before the collapse, and $\lambda$ increases significantly in the collapsed region. Therefore, the strongly magnetized scenario that we discussed has probably not been demonstrated in existing studies, and the field lines are always significantly wrapped in the disk region.

We would like to further bring attention to the RSD in the strong field case by replacing $B_z$ in Equation (\ref{eq:rt_sB}) with the Equation (\ref{eq:lambda}), which yields 
\begin{eqnarray} \label{eq:rt_sB_l}
r_{\rm t} &=& {1\over 4\pi \mu m_{\rm p}} G^{-{1\over2}} k^{-1} \tilde{\lambda} \\
&=& 0.0670~ {\rm AU}\left[ {k \over 8~10^{13} {\rm cm}^{-3}{\rm G}^{-1}}\right]^{-1} \tilde{\lambda}. \nonumber
\end{eqnarray}
This leads to a very small radius compared to the typical values $\sim 20$ AU previously obtained. 
The reason for this apparent contradiction is that $B_z$ can only dominate, with $\lambda$ (\ref{eq:lambda_rI}) satisfying Equation (\ref{eq:lambda_rI_cond_sB}), at small $r$ and thus small $M$, that is, at the early phase of disk formation.
In practice, it is almost impossible to see a RSD dominated by the vertical field component, and the strong field regime is likely irrelevant for disk formation. (see more discussions in Appendix \ref{ap:consistency}).

To estimate the size and mass of the para-RSD upon formation, we assume a very simplified scenario as follows: the very center of the core collapses as if the MB is negligible ($r_{\rm hydro} \propto M$), until it reaches the size where self-regulation through non-ideal MHD effects starts to operate  ($r_{\rm I} \propto \lambda^4/M$). The initial mass and size of the para-RSD can thus be related to the core initial $\lambda$ and $\beta$ by equating Equations (\ref{eq:r_hydro}) and (\ref{eq:rI_sB_wR}), which yields
\begin{eqnarray}\label{eq:Mi}
M_{\rm i} &=& 2^{-{5\over4}} (2\pi)^{-1}  G^{-1} C_{\rm s} \eta_{\rm D}  \beta^{-{1\over4}} \lambda^2\\
&=& 2.67~10^{-4}~M_\odot ~ \left[{ \eta_{\rm D}\over 10^{19} {\rm cm}^2~{\rm s}^{-1}} \right] \left[{\beta \over 0.02}\right]^{-{1\over4}} \lambda^2 \nonumber,~ {\rm and}\\\label{eq:ri}
r_{\rm i} &=& 2^{-{11\over4}} (2\pi)^{-1}  C_{\rm s}^{-1} \eta_{\rm D}  \beta^{{1\over4}} \lambda^2\\
&=& 0.297~{\rm AU} ~ \left[{ \eta_{\rm D}\over 10^{19} {\rm cm}^2~{\rm s}^{-1}} \right] \left[{\beta \over 0.02}\right]^{1\over4} \lambda^2. \nonumber
\end{eqnarray}
The above values should be considered as lower bounds since the non-ideal MHD effects act since the beginning of the collapse and constantly increases $\lambda$. The more efficient the AD in the envelope, the larger the initial disk size.

\subsubsection{Outer partition}
The RSHCS is, on the other hand, a transient product. Combining Equations (\ref{eq:lambda}) and (\ref{eq:rO_sB_wR}) for the strong field case yields
\begin{eqnarray}
\lambda_{\rm O, s} &=& 2 \left[  (2 \pi)^4 G^{3} C_{\rm s}^{-4} \eta_{\rm H}^{-2} M^{3} r_{\rm O}^{-1}  \right]^{1\over8}  \\
&=& 28.2~ \left[{ \eta_{\rm H}\over 10^{19} {\rm cm}^2~{\rm s}^{-1}} \right]^{-{1\over4}}\left[{M \over 0.1 M_\odot}\right]^{3\over8}\left[{r_{\rm O} \over 10 {\rm AU} }\right]^{-{1\over8}}.\nonumber
\end{eqnarray}
Inserting the transport condition for RSHCS existence, Equation (\ref{eq:Hall_vs_Diff}), and the estimated $B_r$, Equation (\ref{eq:Br_ind_T}), yields
\begin{eqnarray}\label{eq:lambda_rO_sB_1}
\lambda_{\rm O, s} &>& 2\left[  (2 \pi)^3  G^2 C_{\rm s}^{-2} |\eta_{\rm H}|^{-2} M^2 \right]^{1\over6} \\
&=& 20.3~  \left[{ |\eta_{\rm H}|\over 10^{19} {\rm cm}^2~{\rm s}^{-1}} \right]^{-{1\over3}} \left[{M \over 0.1 M_\odot}\right]^{1\over3} . \nonumber
\end{eqnarray}
Moreover, applying the strong field presumption, Equation (\ref{eq:cond_sB}), yields the condition
\begin{eqnarray}\label{eq:lambda_sB_BzBphi}
\lambda_{\rm O, s} &>& 2\left[ (2 \pi) G C_{\rm s}^{-1} \eta_{\rm H}^{-1} M \right]^{1\over2} \\
&=& 40.9~  \left[{ \eta_{\rm H}\over 10^{19} {\rm cm}^2~{\rm s}^{-1}} \right]^{-{1\over2}}\left[{M \over 0.1 M_\odot}\right]^{1\over2}. \nonumber
\end{eqnarray}
Both conditions imply that the strong field RSHCS can only exist for low mass, corresponding to relatively early stage.
Moreover, Equation (\ref{eq:lambda_sB_BzBphi}) is likely to be violated first. In consequence, the strong field presumption no longer applies and the weak field regimes should be considered instead.

For the weak field regime, combining Equations (\ref{eq:lambda}) and (\ref{eq:rO_wB_wR}) yields
\begin{eqnarray}
\lambda_{\rm O, w} &=& 2\left[ (2 \pi) G C_{\rm s}^{-1} \eta_{\rm H}^{-1} M \right]^{1\over2} \\
&=& 40.9~  \left[{ \eta_{\rm H}\over 10^{19} {\rm cm}^2~{\rm s}^{-1}} \right]^{-{1\over2}}\left[{M \over 0.1 M_\odot}\right]^{1\over2} \nonumber
\end{eqnarray}
Applying the transport condition for the RSHCS, Equation (\ref{eq:Hall_vs_Diff}), and the estimated $B_r$, Equation (\ref{eq:Br_ind_T}), 
yields
\begin{eqnarray}
\lambda_{\rm O,w} 
&>& 2(2\pi)^{1\over2}  = 5.01,
\nonumber
\end{eqnarray}
which should be easily satisfied even if the initial magnetization is high.

\subsubsection{Ortho-disk}

As for the condition for ortho-disk formation in strong field (small $\lambda$) case, we consider the initial amount of rotation velocity needed to overcome the collapse-induced azimuthal Hall drift, and require that
\begin{eqnarray}\label{eq:uphi_domine_sB}
u_\phi = \sqrt{\beta GM \over r_0} > u_{\rm H, \phi} = {\eta_{\rm H} B_z \over r_0B}.
\end{eqnarray}
Combining Equations (\ref{eq:r0}) and (\ref{eq:uphi_domine_sB}) 
yields
\begin{eqnarray}\label{eq:lambda_sB}
\beta  &>& 2  G^{-2} C_{\rm s}^{2} |\eta_{\rm H}|^{2} M^{-2} \\
&=& 4.5 \times 10^{-4} ~  \left[{ |\eta_{\rm H}|\over 10^{19} {\rm cm}^2~{\rm s}^{-1}}\right] ^{2} \left[{M \over 0.1 M_\odot}\right]^{-2}. \nonumber
\end{eqnarray}
This can be easily satisfied in most conditions, thus an ortho-disk can always form at the beginning of the collapse. 
When solving the same equations for the weak $B_z$ case ($B \approx B_\phi$) along with Equation (\ref{eq:tau_uphi}), the condition for ortho-disk to exist becomes
\begin{eqnarray}\label{eq:lambda_wB}
\beta &>& 256 \pi^2    G^{-2}  C_{\rm s}^{2} |\eta_{\rm H}|^2 M^{-2}\lambda^{-4} \\
&=& 0.57 ~ \left[{ |\eta_{\rm H}| \over 10^{19} {\rm cm}^2~{\rm s}^{-1} }\right]^2 \left[{M \over 0.1 M_\odot }\right]^{-2}  \lambda^{-4}. \nonumber
\end{eqnarray}
Overall, the ortho-disk formation is allowed in more massive or more magnetically supercritical cores. Comparing Equations (\ref{eq:lambda_sB}) and (\ref{eq:lambda_wB}), the amount of rotation needed to form an ortho-disk increases with the magnetization in the weak field regime, and becomes independent of $\lambda$ in the strong field regime.
The two regimes are roughly separated by $\lambda = 6.0$ (here we are referring to the initial value before the collapse). 
All the parallel cases in \citet{Zhao2020a} had initial conditions that allow ortho-disk formation, while the ortho-disk survived only in the very weak field case ($\lambda \sim 9.6$).

In the strong field case,
\begin{eqnarray}
\lambda_{\rm ortho, s} &=& 2\left[  (2 \pi)  G^1 C_{\rm s}^{-2} |\eta_{\rm H}|^{-1} \eta_{\rm D}^{1} M r^{-1} \right]^{1\over2}.
\end{eqnarray}
The strong field presumption (\ref{eq:cond_sB}) requires
\begin{eqnarray}
\lambda_{\rm ortho,s} 
&<& 2(2\pi)^{1\over2}\left[{\eta_{\rm H} \over \eta_{\rm D}}\right]^{1\over2}, 
\end{eqnarray}
and the transport condition 
\begin{eqnarray}\label{eq:Hall_vs_Far}
\tau_{\rm Hall, drift} < \tau_{\rm far}
\end{eqnarray}
reduces to the same form as Equation (\ref{eq:lambda_rO_sB_1}). 
This again means that the ortho-RSD with a dominant vertical field can only exist for small mass. This means the the ortho-disk in the strong field regime could be forced to disappear directly ($M$ decreases), or it can become dominated by the toroidal field due to the loss of magnetic flux before disappearing. In either scenario, a counter-rotating para-RSD develops after the disappearance of the ortho-RSD. Indeed, inserting Equation (\ref{eq:etaD_etaH}) into Equation (\ref{eq:rOrtho_sB}) yields the same estimated $B_z(r_{\rm ortho})$ as in Equation (\ref{eq:BzOrtho_wB}). 
This implies that Equation (\ref{eq:rOrtho_wB_2}) applies to both strong and weak field regimes and there is a smooth transition from the strong field regime to the weak field regime as the ortho-RSD shrinks due to the outward Hall drift.  

In the weak field regime, the transport presumption (\ref{eq:Hall_vs_Far}) 
together with Equations (\ref{eq:lambda}) and (\ref{eq:tau_uphi}) translates into
\begin{eqnarray}\label{eq:lambda_wB_transport}
\lambda &>& 2(2\pi)^{1\over2}   G^{1\over2}C_{\rm s}^{-{1\over2}} |\eta_{\rm H}|^{-{1\over2}} M^{1\over2}  \\
&&=40.9 \left[{ |\eta_{\rm H}| \over 10^{19} {\rm cm}^2~{\rm s}^{-1} }\right]^{-{1\over2}} \left[{M\over 0.1~M_\odot}\right]^{1\over2} \nonumber.
\end{eqnarray}
With sufficiently strong initial parallel rotation and weak magnetization, it is possible that a ortho-RSD forms with $\tau_{\rm far} \simeq \tau_{{\rm Diff, P},z}$ instead. The large radius and large  mass (cf Equations (\ref{eq:Mi}) and (\ref{eq:ri})) make the transport condition (\ref{eq:lambda_wB_transport}) hard to satisfy. If the loss of magnetic flux through radial diffusion and Hall drift fails to increase $\lambda$ quickly enough after the disk formation, the Hall effect will never become dynamically important and the large ortho-RSD survives \citep[see very-weak field case $\lambda \sim 9.6$ in][]{Zhao2020a, Zhao2020b}.

\subsection{Individual non-ideal effects}
\subsubsection{Ambipolar diffusion}
The ambipolar diffusion (AD, or ion-neutral friction) operates at relatively low density and leaks the magnetic field lines outward during the collapse. 
The strength of the AD typically allows the increase of mass-to-flux ratio and determines whether the system is in the strong or weak field case. In the absence of the two other effects, AD regulates the field and $r_{\rm I}$ is found (replacing $\eta_{\rm D}$ by $\eta_{\rm A}$) without an outer RSHCS \citep{Hennebelle2016}. 

\subsubsection{Hall effect}
The Hall effect introduces a specific direction to the rotation-magnetic field configuration. Such effect redistribute the magnetic field lines without necessarily smoothing its distribution. 
This is why a RSD inside a RSHCS is found. 
Our model successfully explained the behavior of the vanishing ortho-disk and the shrinking RSHCS observed by \citet{Zhao2020a}.
In the absence of the Hall effect, only the RSD can exist and some initial rotation is necessary to create the disk. 
We recover the results by \citet{Hennebelle2016}. 

Eventually, in the absence of the two other diffusion mechanisms, the magnetic field can accumulate to a higher strength in the para-disk and drifts upward vertically by Hall advection. 
Such resulting disk will have to satisfy $r_{\rm O}$ and $r_{\rm t}$ 
simultaneously, with very small value of $\eta_{\rm H}$. This means high field strength and small disk size. The para-disk will form at $r_{\rm O}$ and shrink until $r_{\rm t}$ is reached. On the other hand, the edge of the ortho-disk will situate in the region where $\eta_{\rm H}$ is positive, and the field lines drift inward. The field will accumulate until the transition density for $\eta_{\rm H}$ approaches the local density. The whole disk is self-regulated at $\eta_{\rm H} \approx 0$, while it will eventually disappears due to the strong braking. 
Numerical simulations with only Hall effect but no dissipating mechanisms generate very strong local field and thus very small time steps. It is therefore very difficult to evolve such simulations for sufficient long time span and draw reliable conclusions.

\subsubsection{Ohmic dissipation}

The Ohmic dissipation is independent of the field strength and operates at high density, which roughly corresponds to the range where the Hall resistivity becomes positive. 
In the case of the para-disk. 
The field lines drift inward under the Hall effect and accumulates at the disk edge. The azimuthal field is lost through either vertical ($B_\phi$ dominates) or azimuthal ($B_z$ dominates) diffusion. In the absence of Ohmic dissipation, the AD can play the same role, with lower efficiency.

\section{Conclusions}\label{sec:conclusions}
We proposed a model to describe the self-regulated protoplanetary disk size under the complete NIMHD effects. Regimes of different magnetization and rotation strength are discussed. The model predicts a radius of $\sim 10-20$ AU for a wide range of  initial conditions of magnetization and rotation, and also provides the dependencies on the disk mass and magnetic field strength. Our model successfully describes the simulation results of disk formation from previous studies that consider parallel and anti-parallel configurations, including the outer and inner partitions of the para-disk (anti-parallel) and the disappearing small ortho-disk (parallel) \citep{Tsukamoto2015a, Zhao2020a}.
The major conclusions are listed below:
\begin{enumerate}
    \item Protoplanetary disks are small in their early stage. Their size is $\sim 10-20$ AU for a wide range of initial amount of rotation and magnetization. The sub-linear dependencies on combinations of $M$ and $B_z$ or $M$ and $\lambda$ are listed in Table \ref{tab:radius}.
    \item In cases with anti-parallel rotation, no rotation, or weak parallel rotation, a disk (para-disk) forms and rotates in the anti-parallel direction to the initial field. 
    \item The para-disk has two partitions. The outer disk (RSHCS) is short-lived. It flattens and shrinks as the system grows in mass. The inner disk (para-RSD) is long-lived and its size $r\propto M^{1/3}B_z^{-4/9}$.
    \item The strong and weak field regimes are separated by $\lambda \sim 14.3$ in the para-RSD for typical disk parameters, which means that most disks are dominated by the toroidal field.    
    \item A RSHCS dominated by the toroidal field forms when $\lambda \gtrsim 5$, while the vertical field can only dominate when the mass is small.
    \item Rotation in the parallel configuration, if sufficiently strong, forms a short-lived disk (ortho-RSD), with size shrinking in time. 
    \item After the disappearance of the ortho-RSD, a para-RSD will develop due to the Hall effect that torques up the collapsing envelope, in a way very similar to the situation with no initial rotation. 
    \item The para-RSD grows slowly in size as it evolves, while the ortho-RSD shrinks rapidly in size. This could lead to a bimodal size distribution for the observed protoplanetary disks.
    \item The vertical field strength at the edge of the ortho-RSD is roughly $0.01$ G, irrespective of the initial field strength. Due to the outward Hall drift, an ortho-RSD initially dominated by the vertical field could become dominated by the toroidal field. 
    \item If the initial magnetization is extremely weak, the bimodal behavior due to Hall effect is not seen and a large ortho-RSD forms similarly to the para-RSD.
    \item If ortho-RSDs are not observed, it could either be due to the fact that they are short-lived, or that the cores generally have low rotation/high magnetization. This could be potentially used as an observational constraint for prestellar core properties. 
\end{enumerate}

\acknowledgements
The authors thank the anonymous referee for the thorough reading and the suggestions that helped significantly improved the quality of the manuscript.
Y.-N. Lee acknowledges funding from the Ministry of Science and Technology, Taiwan (grant numbers 109-2636-M-003-001 and 110-2124-M-002-012), the grant for Yushan Young Scholar from the Ministry of Education, Taiwan. Pierre Marchand acknowledges the financial support of the Kathryn W. Davis Postdoctoral Fellowship of the American Museum of Natural History. Patrick Hennebelle acknowledges funding from the European Research Council synergy grant ECOGAL (Grant: 855130).

\appendix

\section{Detailed Non-ideal MHD}\label{ap:NIMHD}
We expand the non-ideal terms below in under simplification of axisymmetry. 
The Hall drift velocity describes the drift of field lines along with the charged particles moving inside a magnetic field:
\begin{eqnarray}\label{eq:v_Hall}
\vb*{u}_{\rm H} &=& - {\eta_{\rm H} \over \| \vb*{B} \|} \nabla \times \vb*{B}   \\&=& 
- {\eta_{\rm H} \over \| \vb*{B} \|} 
\left[
 - {\partial B_\phi \over \partial z}\hat{r} 
+\left({\partial B_r \over \partial z} - {\partial B_z \over \partial r} \right)\hat{\phi}
+{\partial rB_\phi \over r\partial r} \hat{z}
\right].  \nonumber
\end{eqnarray}
the AD velocity originate from the magnetic tensor that tends to straighten the field lines and the magnetic pressure that diffuses the field: 
\begin{eqnarray}\label{eq:v_AD}
\vb*{u}_{\rm A} &=& - {\eta_{\rm A} \over \| \vb*{B} \|} \left(\nabla \times \vb*{B}  \right) \times {\vb*{B} \over \| \vb*{B} \|}  
= -{\eta_{\rm A} \over \| \vb*{B} \|^2 }\left[\left(\vb*{B} \cdot \nabla\right) \vb*{B} - {1 \over 2} \nabla \| \vb*{B} \|^2 \right] \\
&=& - {\eta_{\rm A} \over \| \vb*{B} \|^2 }
\left[
\left(B_z{\partial B_r \over \partial z} - B_z{\partial B_z \over \partial r} - {B_\phi \over r}{\partial rB_\phi \over \partial r} \right)\hat{r} 
+\left({B_r \over r}{\partial rB_\phi \over \partial r} +B_z {\partial B_\phi \over \partial z} \right)\hat{\phi}
+\left(B_r{\partial B_z \over \partial r} - B_r{\partial B_r \over \partial z} - B_\phi{\partial B_\phi \over \partial z} \right)\hat{z}
\right], \nonumber
\end{eqnarray}
the Ohmic effect simply diffuses the magnetic field down the gradient slope and smooths the variation of the field strength. This can be regarded as an effective velocity of the field lines \citep{Kunz2009, Braiding2012a, Zhao2020a}, 
which has exactly the same form as $\vb*{u}_{\rm A}$. 
The behavior of the ambipolar diffusion is slightly more complicated while the difference is of limited importance.

The azimuthal velocity in Equation (\ref{eq:momentum}), substituting with the continuity equation, becomes 
\begin{eqnarray}\label{eq:v_phi}
{\partial u_\phi \over \partial t} = -{u_r \over r} {\partial (ru_\phi) \over \partial r} - u_z {\partial u_\phi \over \partial z} + {\partial (rB_rB_\phi) \over \rho r \partial r} + {\partial (B_zB_\phi) \over \rho \partial z},
\end{eqnarray}
where the first two terms correspond to the advection of AM and the last two terms are the acceleration/braking by the magnetic torque.

Equation (\ref{eq:B}) in its expanded form becomes
\begin{eqnarray}
{\partial B_z \over \partial t} &=& {\partial u_zB_r \over \partial r} -{\partial u_rB_z \over \partial r} +{\partial u_{{\rm NI},z}B_r \over \partial r} -{\partial u_{{\rm NI},r}B_z \over \partial r}, \\ 
{\partial B_r \over \partial t} &=& -{\partial u_zB_r \over \partial z} +{\partial u_rB_z \over \partial z} - {\partial u_{{\rm NI},z}B_r \over \partial z} + {\partial u_{{\rm NI},r}B_z \over \partial z}, \label{eq:Br} \\ 
{\partial B_\phi \over \partial t} &=& {\partial u_\phi B_z  \over \partial z}- {\partial u_z B_\phi \over \partial z} - {\partial u_rB_\phi \over \partial r} +{\partial u_\phi B_r\over \partial r} \label{eq:Bphi}
+{\partial u_{{\rm NI},\phi} B_z  \over \partial z}- {\partial u_{{\rm NI},z} B_\phi \over \partial z} - {\partial u_{{\rm NI},r}B_\phi \over \partial r} +{\partial u_{{\rm NI},\phi} B_r\over \partial r}, 
\end{eqnarray}
where the velocity of the charged particles relative to the neutral gas due to the NIMHD effects are summarized as $\vb*{u}_{\rm NI}$. 
The variation of the field strength is subject to different terms of advection $\partial v_j B_i/ \partial j$ and induction by shearing motion $\partial v_i Bj/ \partial j$.

\section{The strength of $B_r$}\label{ap:Br}
It is not straightforward to derive the strength of $B_r$. We therefore present some estimations below by comparing dominant terms in Equation (\ref{eq:Br}).

For the strong field case ($B \approx B_z$), we compare the increase through vertical advection or induction from differential radial velocity to the decrease through radial pressure or tensor diffusion.
There are only two relevant scenario for which an approximated relation can be found between $B_r$ and $B_z$.
If the flow is largely along $z$, $B_r$ is mostly amplified by the vertically converging flow, and lost radially due to the pressure gradient of $B_z$, we obtain
\begin{eqnarray}\label{eq:Br_adv_P}
B_r \approx {\eta_{\rm D} \over u_z r} B_z. 
\end{eqnarray}
This has to satisfy
\begin{eqnarray}
{\tau_{{\rm Diff, P},r} \over \tau_{{\rm Diff, T},r}} = {\eta_{\rm D} \over h u_z} < 1,
\end{eqnarray}
which stays true for large $h$.
In contrast, when the flow is misaligned with the field, the induction from vertical shear bends the field lines, which tend to straighten themselves, we obtain 
\begin{eqnarray}\label{eq:Br_ind_T}
B_r \approx {u_r h \over \eta_{\rm D}} B_z.
\end{eqnarray}
This time 
\begin{eqnarray}
{\tau_{{\rm Diff, P},r} \over \tau_{{\rm Diff, T},r}} = {ru_r \over \eta_{\rm D}} >1
\end{eqnarray}
is satisfied for large $r$.
In general, tensor diffusion dominates near the midplane while pressure diffusion is relevant for higher latitude, and the separation between the two regions depends on competition between the diffusion and collapse.

In the strongly collapsing regime (large $\lambda$), it is not straightforward to deduce whether $B_r$ or $B_\phi$ is the dominant component. 
We therefore look for solutions under both conditions and check if the assumptions are violated. 
Assume that the radial field component reaches an equilibrium when the radial drag is compensated by the vertical pressure diffusion, $B_r$ follows the same expression as in Equation (\ref{eq:Br_ind_T}).
If we assume 
\begin{eqnarray}\label{eq:Br_Bphi}
B \approx B_r, 
\end{eqnarray}
solving the Equation (\ref{eq:tau_RSHCS}) for the RSHCS in this regime yields
\begin{eqnarray}\label{eq:rOrtho_Br}
r_{\rm O} &=&\left[ (2\pi)^{-4}G^{1}C_{\rm s}^{2} |\eta_{\rm H}|^{2} \eta_{\rm A}^{2} M^{5}B_z^{-8}\right]^{1\over17}\nonumber\\
&=&  14.5 ~ {\rm AU} ~  \left[{|\eta_{\rm H}| \over 10^{19} {\rm cm}^2~{\rm s}^{-1}}\right]^{2\over17} \left[{\eta_{\rm A} \over 10^{19} {\rm cm}^2~{\rm s}^{-1}}\right]^{2\over17}\left[{M \over 0.1 M_\odot}\right]^{5\over17} \left[{B_z \over 0.1 {\rm G}}\right]^{-{8\over17}}.
\end{eqnarray}

Here we apply the ambipolar diffusion since we are discussing about the balance in the inner envelope. 
The solution of $r_{\rm O}$ in Equation (\ref{eq:rOrtho_Br}) should satisfy the inequality
\begin{eqnarray}\label{eq:Br_Bphi}
B_r > B_\phi, 
\end{eqnarray}
which then yields
\begin{eqnarray}
r_{\rm O} >  G C_{\rm s}^{-2} |\eta_{\rm H}|^{-2} \eta_{\rm A}^{2} M 
= 2223 ~ {\rm AU} ~  \left[{\eta_{\rm A} \over |\eta_{\rm H}|}\right]^{2}\left[{M \over 0.1 M_\odot}\right],
\end{eqnarray}
which is in obvious contradiction. 
Unless $|\eta_{\rm H}|$ is significantly larger than $\eta_{\rm A}$ in some rare cases, then Equation (\ref{eq:rOrtho_Br}) is the relevant solution and (\ref{eq:Br_Bphi}) is satisfied.

\section{Self-consistency check}\label{ap:consistency}
For the solutions to be self-consistent, we check if the presumptions are satisfied.
For the RSHCS, the inward radial Hall drift must dominate over the outward AD. 
On the other hand, the azimuthal/vertical Ohmic dissipation must dominate over the inward Hall drift at the para-RSD edge. 
As for ortho-RSD formation, the outward Hall drift  dominates over the Faraday induction.
We also check if the relative strength among $B_z$, $B_\phi$, and $B_r$ is consistent  with the solutions. The results are listed in Table \ref{tab:condition}. 
The expressions of the equilibrium radius are used to eliminate the magnetic terms in the inequalities. The condition can then be re-written in terms of a size limit as function of mass.

Most presumptions are easily satisfied, confirming the validity of the model predictions. 
Interestingly, some conditions require closer inspection and particular behaviors can be identified:
\begin{enumerate}
   \item The RSHCS must become dominated by the toroidal field as the disk grows to $rM  \gtrsim0.05~ AU~M_\odot$.
   \item As an initially strongly magnetized para-RSD grows in size, the field will eventually becomes rotationally winded and $B_\phi$ will become larger than $B_z$.  
   \item The strong field presumption for the ortho-RSD is quickly violated as the disk grows in mass. The relevant description then becomes the weak field ortho-RSD.
\end{enumerate}

\begin{deluxetable}{ccccch}
\tablecaption{Presumptions for the model simplification\label{tab:condition}}
\tablehead{
\colhead{Type} & \colhead{$B$ field}  & \colhead{Presumption} & \colhead{$r$ limit} & \colhead{Reference value (AU)} & \nocolhead{Limit (AU, $10^{19} {\rm cm}^2~{\rm s}^{-1}$, $0.1~M_\odot$)} 
}
\decimals
\startdata
RSHCS & $B_z$  & $\tau_{\rm Hall, drift} < \tau_{{\rm Diff, T}, r}$ &   $r <  G^{1\over3}C_{\rm s}^{-{4\over 3}} |\eta_{\rm H}|^{{2\over3}}  M^{1\over3}$ & 135 &$r <135   ~ \delta_\phi^{-2/3}  |\eta_{\rm H}|^{-2/3} \eta_{\rm O}^{4/3} M^{1/3}$ \\
RSHCS & $B_z$  & $B_z > B_\phi$ &$r<  G^{-1} |\eta_{\rm H}|^{2}  M^{-1}$& 0.500 & $r< 0.5 ~ \delta_\phi^{-2}  |\eta_{\rm H}|^{2}  M^{-1}$ \\
para-RSD &  $B_z$  & $\tau_{\rm Hall, drift} > \tau_{{\rm Diff, T}, \phi}$ &  $r <  G^{1\over3}C_{\rm s}^{-{4\over 3}} |\eta_{\rm H}|^{-{2\over3}} \eta_{\rm D}^{4\over3} M^{1\over3}$ & 135 &$r <135   ~ \delta_\phi^{-2/3}  |\eta_{\rm H}|^{-2/3} \eta_{\rm D}^{4/3} M^{1/3}$ \\
para-RSD &  $B_z$  & $B_z > B_\phi$ &  $r <   C_{\rm s}^{-1}  \eta_{\rm D} $ & 33 &  $r < 33 ~   \delta_\phi^{-1}  \eta_{\rm O} $ \\
ortho-RSD &  $B_z$  & $\tau_{\rm Hall, drift} < \tau_{\rm far}$ &  $r <  G^{1\over3}C_{\rm s}^{-{4\over 3}} |\eta_{\rm H}|^{-{2\over3}} \eta_{\rm D}^{{4\over3}}  M^{1\over3}$ & 135 &$r <135   ~ \delta_\phi^{-2/3}  |\eta_{\rm H}|^{-2/3} \eta_{\rm O}^{4/3} M^{1/3}$ \\
ortho-RSD &  $B_z$  & $B_z > B_\phi$ &  $r > GC_{\rm s}^{-2} |\eta_{\rm H}|^{-2} \eta_{\rm D}^{2}  M$ &2223&  $r > 2223 ~|\eta_{\rm H}|^{-2} \eta_{\rm O}^{2} M$ \\
\hline
RSHCS & $B_\phi$  & $\tau_{\rm Hall, drift} < \tau_{{\rm Diff, T}, r}$ &  $M >  G^{-1} C_{\rm s}  |\eta_{\rm H}|$ & 0.0015 ($M_\odot$) &  $M > 0.0015 ~ \delta_\phi^{-2} \delta_r   |\eta_{\rm H}|$ \\
RSHCS & $B_\phi$  & $B_\phi > B_z$ & $r > G^{-1} |\eta_{\rm H}|^{2}  M^{-1}$ &0.500 &$r > 0.500 ~   |\eta_{\rm H}|^{2}  M^{-1}$ \\
RSHCS & $B_\phi$  & $ B_\phi>B_r$ &   $r <  GC_{\rm s}^{-2} |\eta_{\rm H}|^{-2} \eta_{\rm D}^{2} M$ &2223&  $r > 2223 ~|\eta_{\rm H}|^{-2} \eta_{\rm O}^{2} M$ \\
para-RSD &  $B_\phi$  & $\tau_{\rm Hall, drift} > \tau_{{\rm Diff, P},z}$ &  $r <  G C_{\rm s}^{-2} |\eta_{\rm H}|^{-2} \eta_{\rm D}^{2} M$ & 2223 & $r < 2223   ~   |\eta_{\rm H}|^{-2} \eta_{\rm D}^{2} M$\\
para-RSD &  $B_\phi$  & $B_\phi > B_z$ &  $r > ~    C_{\rm s}^{-1} \eta_{\rm D} $ &33&  $r > 33 ~   \delta_\phi^{-1}  \eta_{\rm D} $ \\
ortho-RSD &  $B_\phi$  & $\tau_{\rm Hall, drift} < \tau_{\rm far}$ &  $r < (2\pi)^{-{1\over4}}  (k\mu m_{\rm p})^{-{1\over2}}(n_{\rm M}/n_{\rm e})^{{1\over4}} C_{\rm s}^{-{3\over 4}} |\eta_{\rm H}|^{{1\over4}} M^{{1\over4}} $ & 30.8 &.\\
ortho-RSD &  $B_\phi$  & $ B_\phi > B_z$ &  $r < (2\pi)^{-{2\over7}} (k\mu m_{\rm p})^{-{4\over7}}(n_{\rm M}/n_{\rm e})^{{2\over7}} G^{{1\over7}} C_{\rm s}^{-{6\over 7}}  M^{{3\over7}}  $ & 55.5 &.\\
\enddata
\tablecomments{The numerical reference values are calculated with $\eta= 10^{19} {\rm cm}^2{\rm s}^{-1}$ and $M=0.1~M_\odot$.}
\end{deluxetable}

\section{Transition density of the Hall resistivity }\label{ap:rho_crit}
At low density, the dynamics of the magnetic field is dominated by the positively charged ions, while the electrons start to dominate at high densities. Since the Hall resistivity depends on both the particle density and the field strength, the density at which this transition happens is a function of the field strength. Following the definitions in \citet[][see details therein]{Marchand2016}, the Ohmic, Hall, and ambipolar resistivities are
\begin{eqnarray}
\eta_{\rm O} = {1\over \sigma_{||}},~
\eta_{\rm H} =  {\sigma_{\rm H} \over \sigma_\perp^2 + \sigma_{\rm H}^2}, ~~{\rm and}~~
\eta_{\rm A} =  {\sigma_\perp \over \sigma_\perp^2 + \sigma_{\rm H}^2} - {1\over \sigma_{||}}
\end{eqnarray}
where
\begin{eqnarray}
\sigma_{||} = \sum_i \sigma_i,~
\sigma_\perp = \sum_i {\sigma_i \over 1+(\omega_i \tau_{i{\rm n}})^2} ~~{\rm and}~~
\sigma_{\rm H} = -\sum_i {\sigma_i \omega_i \tau_{i{\rm n}} \over 1+(\omega_i \tau_{i{\rm n}})^2}, 
\end{eqnarray}
where the subscript $i$ stands for all charged particle species and n for neutrals,
with $\sigma_i = n_iq_i^2\tau_{i{\rm n}}/m_i$, $\omega_i = q_iB/(m_ic)$, and $\tau_{i{\rm n}} = (1+m_i/m_{{\rm H}_2})/(a_{i{\rm He}}n_{{\rm H}_2}\langle\sigma_{\rm coll}w\rangle_i)$.
The sign inversion of $\eta_{\rm H}$ happens at the transition between the dominance of singly charged positive ions and that of electrons, which are the two most abundant species in the density range of interest. 
Given a low cyclotron frequency for ions and a large one for electrons, we have $\omega_{\rm M} \tau_{{\rm Mn}} \ll $ and $\omega_{\rm e} \tau_{{\rm en}} \gg 1$, where M and e stand for ions and electrons, respectively. 

\begin{eqnarray}\label{eq:etaH_etaD}
{\eta_{\rm H} \over \eta_{\rm A} + \eta_{\rm O}} = {\sigma_{\rm H} \over \sigma_\perp} \approx
{{{\sigma_{\rm M} \omega_{\rm M} \tau_{\rm Mn}} \over 1+(\omega_{\rm M} \tau_{{\rm Mn}})^2} + {\sigma_{\rm e} \omega_{\rm e} \tau_{\rm en} \over 1+(\omega_{\rm e} \tau_{\rm en})^2} \over
{\sigma_{\rm M}  \over 1+(\omega_{\rm M} \tau_{{\rm Mn}})^2} + {\sigma_{\rm e} \over 1+(\omega_{\rm e} \tau_{\rm en})^2} } 
\approx { {\sigma_{\rm M} \omega_{\rm M} \tau_{\rm Mn}} + {\sigma_{\rm e} \over \omega_{\rm e} \tau_{{\rm en}}} \over  {\sigma_{\rm M} } + {\sigma_{\rm e} \over \omega_{\rm e}^2 \tau_{{\rm en}}^2} } 
\approx
\left({n_{{\rm H}_2}\over n_{\rm t}}-{n_{\rm t} \over n_{{\rm H}_2}}\right) \left(n_{\rm e} \over n_{\rm M}\right)^{1 \over 2},
\end{eqnarray}
where
\begin{eqnarray}
n_{\rm t} = {e \over c m_{{\rm H}_2} a_{{\rm M}{\rm He}} \langle\sigma_{\rm coll}w\rangle_{\rm M}} \left( n_{\rm M} \over n_{\rm e} \right)^{1\over 2} B = kB.
\end{eqnarray}
This conditions $\eta_{\rm H} =0 $ is satisfied when
\begin{eqnarray}
n \approx n_{{\rm H}_2} = n_{\rm t}.
\end{eqnarray}
Applying $n_{\rm M}/n_{\rm e} \approx 107$ \citep{Marchand2021} and $\langle\sigma_{\rm coll}w\rangle_{\rm M} \approx 2.4 \times 10^{-9} ({\rm cm^3~s}^{-1}) C_{\rm s}({\rm km~s}^{-1})^{0.6}  \approx 9.1~10^{-10}$ \citep{Pinto2008}, we obtain $k \approx 5.2 \times 10^{13} {\rm cm}^{-3}{\rm G}^{-1}$, which is not far from the value measured directly from simulations.

\bibliography{bib_nimhd}

\begin{thebibliography}{}
\expandafter\ifx\csname natexlab\endcsname\relax\def\natexlab#1{#1}\fi
\providecommand{\url}[1]{\href{#1}{#1}}
\providecommand{\dodoi}[1]{doi:~\href{http://doi.org/#1}{\nolinkurl{#1}}}
\providecommand{\doeprint}[1]{\href{http://ascl.net/#1}{\nolinkurl{http://ascl.net/#1}}}
\providecommand{\doarXiv}[1]{\href{https://arxiv.org/abs/#1}{\nolinkurl{https://arxiv.org/abs/#1}}}

\bibitem[{{Andrews}(2020)}]{Andrews2020}
{Andrews}, S.~M. 2020, \araa, 58, 483,
  \dodoi{10.1146/annurev-astro-031220-010302}

\bibitem[{{Andrews} {et~al.}(2018){Andrews}, {Huang}, {P{\'e}rez}, {Isella},
  {Dullemond}, {Kurtovic}, {Guzm{\'a}n}, {Carpenter}, {Wilner}, {Zhang}, {Zhu},
  {Birnstiel}, {Bai}, {Benisty}, {Hughes}, {{\"O}berg}, \&
  {Ricci}}]{Andrews2018}
{Andrews}, S.~M., {Huang}, J., {P{\'e}rez}, L.~M., {et~al.} 2018, \apjl, 869,
  L41, \dodoi{10.3847/2041-8213/aaf741}

\bibitem[{{Belloche}(2013)}]{Belloche2013}
{Belloche}, A. 2013, in EAS Publications Series, Vol.~62, EAS Publications
  Series, ed. P.~{Hennebelle} \& C.~{Charbonnel}, 25--66,
  \dodoi{10.1051/eas/1362002}

\bibitem[{{Braiding} \& {Wardle}(2012)}]{Braiding2012a}
{Braiding}, C.~R., \& {Wardle}, M. 2012, \mnras, 422, 261,
  \dodoi{10.1111/j.1365-2966.2012.20601.x}

\bibitem[{{Gaudel} {et~al.}(2020){Gaudel}, {Maury}, {Belloche}, {Maret},
  {Andr{\'e}}, {Hennebelle}, {Galametz}, {Testi}, {Cabrit}, {Palmeirim},
  {Ladjelate}, {Codella}, \& {Podio}}]{Gaudel2020}
{Gaudel}, M., {Maury}, A.~J., {Belloche}, A., {et~al.} 2020, \aap, 637, A92,
  \dodoi{10.1051/0004-6361/201936364}

\bibitem[{{Goodman} {et~al.}(1993){Goodman}, {Benson}, {Fuller}, \&
  {Myers}}]{Goodman1993}
{Goodman}, A.~A., {Benson}, P.~J., {Fuller}, G.~A., \& {Myers}, P.~C. 1993,
  \apj, 406, 528, \dodoi{10.1086/172465}

\bibitem[{{Hennebelle} {et~al.}(2016){Hennebelle}, {Commer{\c{c}}on},
  {Chabrier}, \& {Marchand}}]{Hennebelle2016}
{Hennebelle}, P., {Commer{\c{c}}on}, B., {Chabrier}, G., \& {Marchand}, P.
  2016, \apjl, 830, L8, \dodoi{10.3847/2041-8205/830/1/L8}

\bibitem[{{Hennebelle} {et~al.}(2020){Hennebelle}, {Commer{\c{c}}on}, {Lee}, \&
  {Chabrier}}]{Hennebelle2020b}
{Hennebelle}, P., {Commer{\c{c}}on}, B., {Lee}, Y.-N., \& {Chabrier}, G. 2020,
  \apj, 904, 194, \dodoi{10.3847/1538-4357/abbfab}

\bibitem[{{Inutsuka} {et~al.}(2010){Inutsuka}, {Machida}, \&
  {Matsumoto}}]{Inutsuka2010}
{Inutsuka}, S.-i., {Machida}, M.~N., \& {Matsumoto}, T. 2010, \apjl, 718, L58,
  \dodoi{10.1088/2041-8205/718/2/L58}

\bibitem[{{Kunz} \& {Mouschovias}(2009)}]{Kunz2009}
{Kunz}, M.~W., \& {Mouschovias}, T.~C. 2009, \mnras, 399, L94,
  \dodoi{10.1111/j.1745-3933.2009.00731.x}

\bibitem[{{Lee} {et~al.}(2021){Lee}, {Charnoz}, \& {Hennebelle}}]{Lee2021}
{Lee}, Y.-N., {Charnoz}, S., \& {Hennebelle}, P. 2021, arXiv e-prints,
  arXiv:2102.07963.
\newblock \doarXiv{2102.07963}

\bibitem[{{Marchand} {et~al.}(2018){Marchand}, {Commer{\c{c}}on}, \&
  {Chabrier}}]{Marchand2018}
{Marchand}, P., {Commer{\c{c}}on}, B., \& {Chabrier}, G. 2018, \aap, 619, A37,
  \dodoi{10.1051/0004-6361/201832907}

\bibitem[{{Marchand} {et~al.}(2021){Marchand}, {Guillet}, {Lebreuilly}, \& {Mac
  Low}}]{Marchand2021}
{Marchand}, P., {Guillet}, V., {Lebreuilly}, U., \& {Mac Low}, M.~M. 2021,
  \aap, 649, A50, \dodoi{10.1051/0004-6361/202040077}

\bibitem[{{Marchand} {et~al.}(2016){Marchand}, {Masson}, {Chabrier},
  {Hennebelle}, {Commer{\c{c}}on}, \& {Vaytet}}]{Marchand2016}
{Marchand}, P., {Masson}, J., {Chabrier}, G., {et~al.} 2016, \aap, 592, A18,
  \dodoi{10.1051/0004-6361/201526780}

\bibitem[{{Marchand} {et~al.}(2019){Marchand}, {Tomida}, {Commer{\c{c}}on}, \&
  {Chabrier}}]{Marchand2019}
{Marchand}, P., {Tomida}, K., {Commer{\c{c}}on}, B., \& {Chabrier}, G. 2019,
  \aap, 631, A66, \dodoi{10.1051/0004-6361/201936215}

\bibitem[{{Marchand} {et~al.}(2020){Marchand}, {Tomida}, {Tanaka},
  {Commer{\c{c}}on}, \& {Chabrier}}]{Marchand2020}
{Marchand}, P., {Tomida}, K., {Tanaka}, K. E.~I., {Commer{\c{c}}on}, B., \&
  {Chabrier}, G. 2020, \apj, 900, 180, \dodoi{10.3847/1538-4357/abad99}

\bibitem[{{Masson} {et~al.}(2016){Masson}, {Chabrier}, {Hennebelle}, {Vaytet},
  \& {Commer{\c{c}}on}}]{Masson2016}
{Masson}, J., {Chabrier}, G., {Hennebelle}, P., {Vaytet}, N., \&
  {Commer{\c{c}}on}, B. 2016, \aap, 587, A32,
  \dodoi{10.1051/0004-6361/201526371}

\bibitem[{{Maury} {et~al.}(2018){Maury}, {Girart}, {Zhang}, {Hennebelle},
  {Keto}, {Rao}, {Lai}, {Ohashi}, \& {Galametz}}]{Maury2018}
{Maury}, A.~J., {Girart}, J.~M., {Zhang}, Q., {et~al.} 2018, \mnras, 477, 2760,
  \dodoi{10.1093/mnras/sty574}

\bibitem[{{Maury} {et~al.}(2019){Maury}, {Andr{\'e}}, {Testi}, {Maret},
  {Belloche}, {Hennebelle}, {Cabrit}, {Codella}, {Gueth}, {Podio}, {Anderl},
  {Bacmann}, {Bontemps}, {Gaudel}, {Ladjelate}, {Lef{\`e}vre}, {Tabone}, \&
  {Lefloch}}]{Maury2019}
{Maury}, A.~J., {Andr{\'e}}, P., {Testi}, L., {et~al.} 2019, \aap, 621, A76,
  \dodoi{10.1051/0004-6361/201833537}

\bibitem[{{Pinto} \& {Galli}(2008)}]{Pinto2008}
{Pinto}, C., \& {Galli}, D. 2008, \aap, 484, 17,
  \dodoi{10.1051/0004-6361:20078819}

\bibitem[{{Shu}(1977)}]{Shu1977}
{Shu}, F.~H. 1977, \apj, 214, 488, \dodoi{10.1086/155274}

\bibitem[{{Tsukamoto} {et~al.}(2015{\natexlab{a}}){Tsukamoto}, {Iwasaki},
  {Okuzumi}, {Machida}, \& {Inutsuka}}]{Tsukamoto2015a}
{Tsukamoto}, Y., {Iwasaki}, K., {Okuzumi}, S., {Machida}, M.~N., \& {Inutsuka},
  S. 2015{\natexlab{a}}, \apjl, 810, L26, \dodoi{10.1088/2041-8205/810/2/L26}

\bibitem[{{Tsukamoto} {et~al.}(2015{\natexlab{b}}){Tsukamoto}, {Iwasaki},
  {Okuzumi}, {Machida}, \& {Inutsuka}}]{Tsukamoto2015b}
---. 2015{\natexlab{b}}, \mnras, 452, 278, \dodoi{10.1093/mnras/stv1290}

\bibitem[{{Wurster} {et~al.}(2014){Wurster}, {Price}, \&
  {Ayliffe}}]{Wurster2014}
{Wurster}, J., {Price}, D., \& {Ayliffe}, B. 2014, \mnras, 444, 1104,
  \dodoi{10.1093/mnras/stu1524}

\bibitem[{{Wurster} {et~al.}(2016){Wurster}, {Price}, \& {Bate}}]{Wurster2016}
{Wurster}, J., {Price}, D.~J., \& {Bate}, M.~R. 2016, \mnras, 457, 1037,
  \dodoi{10.1093/mnras/stw013}

\bibitem[{{Zhao} {et~al.}(2020{\natexlab{a}}){Zhao}, {Caselli}, {Li},
  {Krasnopolsky}, {Shang}, \& {Lam}}]{Zhao2020a}
{Zhao}, B., {Caselli}, P., {Li}, Z.-Y., {et~al.} 2020{\natexlab{a}}, \mnras,
  492, 3375, \dodoi{10.1093/mnras/staa041}

\bibitem[{{Zhao} {et~al.}(2020{\natexlab{b}}){Zhao}, {Caselli}, {Li},
  {Krasnopolsky}, {Shang}, \& {Lam}}]{Zhao2020b}
---. 2020{\natexlab{b}}, arXiv e-prints, arXiv:2009.07820.
\newblock \doarXiv{2009.07820}

\end{thebibliography}
\bibliographystyle{aasjournal}



\end{CJK*}
\end{document}